\documentclass[fleqn,usenatbib,useAMS]{mnras}

\usepackage[T1]{fontenc}

\usepackage{graphicx}	
\usepackage{amsmath}	
\usepackage{amssymb}	
\usepackage{multicol}        
\usepackage{bm}		
\usepackage{pdflscape}	
\usepackage{color}
\usepackage[dvipsnames]{xcolor}
\usepackage{multirow}
\usepackage{ulem}
\usepackage{ae,aecompl}
\usepackage{newtxtext,newtxmath}

\newcommand{\OI}{O\,{\sevensize I}\ }

\newcommand{\HeI}{He\,{\sevensize I}\ }

\newcommand{\kms}{\,km\,s$^{-1}$} 

\newcommand{\Msun}{\ensuremath{\rm M_\odot}}

\newcommand{\Lsun}{\ensuremath{\rm L_\odot}}
\newcommand{\Rsun}{\ensuremath{\rm R_\odot}}

\newcommand{\HDtwentyfive}{HD\,253659}
\newcommand{\hdtwentyfive}{HD\,253659}

\newcommand{\tess}{\textit{TESS}}

\newcommand{\js}{\textcolor{violet}}

 \title[Eruptive Be stars PY\,Gem and HD\,253659]{Central stars of newly discovered infrared nebulae: eruptive Be stars PY\,Gem and HD\,253659} 

\author[O. V. Maryeva et al.]{Olga~Maryeva,$^{1}$\thanks{E-mail: olga.maryeva@asu.cas.cz}
          Julieta~S\'{a}nchez Arias$,^{1}$ Sergey~Karpov,$^{2}$ Michaela~Kraus,$^{1}$  Michalis~Kourniotis,$^{1}$ \\
\\          
{\LARGE \rm  and Sabina Mammadova$^{3}$ }  \\    
\\
$^{1}$ Astronomical Institute of the Czech Academy of Sciences, Fri\v{c}ova 298, 25165 Ond\v{r}ejov, Czech Republic \\
$^{2}$ Institute of Physics of the  Czech Academy of Sciences, CZ-182 21 Prague 8, Czech Republic \\
$^{3}$ Shamakhy Astrophysical Observatory, Y.Mammadaliyev, AZ5626, Azerbaijan 
}
\date{} 

\pubyear{2026}

\begin{document}
\label{firstpage}
\pagerange{\pageref{firstpage}--\pageref{lastpage}}
\maketitle
\begin{abstract}
 The presence of a nebula around massive hot stars often works as an indicator that the object is either in an advanced evolutionary stage or a rare product of close binary interaction.  Here, we focus on two Be stars  -- PY\,Gem and \HDtwentyfive\ -- whose nebulae were detected with the {\it Wide-field Infrared Survey Explorer}. We estimate their basic physical parameters from modelling the spectral energy distribution and  examine their variability. We combine archival photometric data from several ground-based survey telescopes and from the {\it Transiting Exoplanet Survey Satellite} (\tess) with our own data from dedicated multi-colour photometric and spectroscopic monitoring, and conclude that,  
despite having similar stellar properties, the photometric variability of the two objects is strikingly different. For \HDtwentyfive\ we detect continuous eruptive variability and a major outburst that took place between 2016 and 2022 along with numerous flicker events seen in the \tess\ data.  \HDtwentyfive\ also exhibits loop-like behaviour in the colour--magnitude diagram, consistent with build-up and dissipation phases of a circumstellar disk seen close to pole-on. In contrast, during the past $\sim 20$ years PY\,Gem has lost its eruptive variability and shows only small-amplitude $p$ and $g$-mode pulsations classifying PY\,Gem as $\beta$~Cephei hybrid pulsator. An incoherent low-frequency signal is identified as a \v{S}tefl frequency, which seems to be supported by the cyclic variation of the emission-line profiles. Detailed analysis of these stars indicates that both objects are inconsistent with evolved massive stars, and that the origin of their nebulae is more likely linked to the physics of the Be phenomenon.
\end{abstract}

\section{Introduction}\label{sec:intro}

Circumstellar nebulae and shells may be found around a variety of astrophysical objects. The formation of nebulae can have different origins but, in most cases, it is linked to mass ejections at late stages of stellar evolution, and various kinds of interaction in binary systems, such as wind collision, Roche-lobe overflow or merging. The ejected matter typically expands and dilutes so that nebulae are only detectable during a certain amount of time. Their material is ionized by the radiation of the central star and the nebulae can be seen in recombination lines (such as the hydrogen series) and classical nebular lines, such as forbidden emission lines of ionized and neutral metals. Dust can form in cool nebulae, and its thermal emission can be observed at infrared wavelengths. 

In contrast to nebulae, shells often result from the interaction between stellar winds and the surrounding interstellar medium (ISM). Such interactions often generate shocks in which the material of both the wind and the ISM is compressed and heated. This may result in impressive shells and arc-like structures,
such as the wind-blown bubbles around massive stars \citep[e.g.][and references therein]{2023Galax..11...78D} and the bowshocks around OB-type runaway stars \citep[e.g.][]{2012A&A...538A.108P, 2015A&A...578A..45P, 2017AJ....154..201K}.

\begin{table*}\centering
\caption{Stars studied in this work. The photometry is taken from the Gaia DR3 Synthetic Photometry catalogue \citep{gaiadr3syn}. Variability type is according to the AAVSO International Variable Stars Index \citep{VariableStarVSX2006}. Spectral types are  from:
$^{a}$~{\citet{Guetter1968}}, $^{b}$~{\citet{Morgan1955}}. 
The distance to the objects is taken from \citet{Bailer-Jones2021}, $\mu$ is the proper motion from \citet{GaiaEDR3}. The size corresponds to the full extent of the nebula on the sky along the longest axis. }
\label{tab:stars} 
{\begin{tabular}{l lc ccc ccc c}
\hline
\multicolumn{1}{c}{\multirow{2}{*}{Star}} 
&  \multicolumn{1}{c}{\multirow{2}{*}{Sp. type}} & Var.& Distance & $\mu$          & \multicolumn{1}{c}{\multirow{2}{*}{RUWE}}         & $B$  &  $V$& $R$ & Size \\
    &                                               & type& [pc] & [mas yr$^{-1}$]&  &  [mag]&[mag]&[mag]  & [pc] \\ 
\hline
PY\,Gem         & B1~Vne$^{a}$ & $\gamma$\,Cas & $1401_{-63}^{+48}$  & 1.301 & 0.977 & 8.52  & 8.46 & 8.37 & 1.83\\ 
\HDtwentyfive   &  B0.5~V:nne$^{b}$  & $\gamma$\,Cas  & $1698_{-58}^{+66}$  & 1.002 & 1.106 & 9.59 & 9.16 &   8.86 & 1.48 \\ 
\hline
\end{tabular}}
\end{table*}

Significant progress in the detection of new circumstellar nebulae was achieved with data from the {\it Spitzer} Space Telescope. Based on these multi-band images, several groups independently discovered many dozens of circumstellar shells \citep{GKF2010, 2010AJ....139.1542M, Wachter2010nebulae}. Other space missions, such as the {\it Midcourse Space Experiment (MSX)} and the {\it Wide-field Infrared Survey Explorer (WISE)}, provided also useful images based on which nebulae around massive stars were detected, but the spatial resolution of these missions was significantly lower, hiding many structural details of the nebulae. Nevertheless, follow-up spectroscopy of the central stars of
many of these shells led to the discovery of dozens of Wolf--Rayet (WR) stars, luminous
blue variables (LBVs), and other massive stars \citep{Clark2003, Gvaramadze2009WR, GKF2010, Gvaramadze2010,  Mauerhan2010WR, Wachter2010nebulae, StringfellowGvaramadze2012, 2020AJ....160..166C} and the lists of Galactic bona fide and candidate LBVs have been significantly expanded \citep{Kniazev2015LBV, Kniazev2016, Richardson2018}.

 We continue to explore central stars of circumstellar nebulae and shells, a list of which is partially published  in \citet{Gvaramadze2010} and \citet{Kniazev2016}. In the current work, we analyze Be stars found in the sample,  as 
circumstellar nebulae are not a typical feature of classical Be stars. Moreover, their presence is difficult to reconcile with the standard evolutionary picture, since main-sequence stars are not generally expected to undergo episodes of sufficiently strong mass loss to produce extended circumstellar nebulae.  
The discovery of nebulae around rapidly rotating stars at or near the end of the main sequence phase \citep[and references therein]{Gvaramadze2018} supported
the hypothesis of Lamers et al. (2001) that the first significant mass loss in rotating stars can occur even before the end of the core hydrogen burning stage. In this context, Be stars, which are rapidly rotating main-sequence B-type stars, are particularly interesting because they 
have phases during which they are surrounded by ionized gaseous disks, most likely formed by  their rapid rotation \citep[e.g.][]{2003PASP..115.1153P}, which by itself may lead already to mechanical mass-loss when reaching the critical value \citep{2011A&A...527A..84K, 2013A&A...553A..25G}. Or, mass ejections into (quasi-)Keplerian orbits may be triggered by rapid rotation paired with non-radial pulsations of the central star \citep{2003A&A...411..229R,RiviniusCarciofi2013} that can cause discrete mass ejections known as "flickers" (when occurring on short timescales) or "outbursts"   \citep[e.g.,][]{2025Galax..13...77C}. Parts of the ejected (or disk) matter will certainly dissipate or interact with the ISM. On the other hand, recent simulations of Be binary systems have shown that material from the Be star disk  can also end up revolving on circumbinary orbits \citep{2025A&A...698A.309R}, which might form the base of a nebula around such systems.

 We selected two objects -- PY\,Gem and \HDtwentyfive\ -- that are plausible candidates for being associated with the discovered nebula. Both are visible in the northern hemisphere  with brightness  $V<10$~mag, which allows us to study them using medium and high-resolution spectroscopy and several photometric survey facilities.  The nebulae were not revealed in the large-scale study of infrared nebulae around bright massive stars \citep{Bodensteiner2018}, as the luminosities of their central stars are outside the magnitude limit in that study (which required a magnitude in $V$ of less than 6.5), and they are not mentioned in other publications. Both stars are listed as eruptive variable stars of $\gamma$\,Cas type ($\gamma$\,Cas variables) in the International Variable Star Index\footnote{The General Catalog of Variable Stars (GCVS) \citep{Samus2017} only lists PY\,Gem as a Be star, while \HDtwentyfive\ is absent there.} \citep{VariableStarVSX2006}. 
There are no X-ray sources associated with these stars, and therefore they are not included in the list of $\gamma$\,Cas analogs\footnote{By definition in GCVS, $\gamma$\,Cas variables are stars that exhibit long-term photometric variability associated with the formation  and dissipation of the disk. The similarly sounding but unrelated term ``$\gamma$\,Cas analogs'', which is outside the formal GCVS classification scheme, is becoming increasingly common to refer to a specific type of X-ray behaviour \citep[see][and references therein]{Rivinius2026}.} by \citet{Naze2018Xray-Oe-Be}.

Table~\ref{tab:stars} lists spectral classification of the stars taken from the literature and  their photometric magnitudes according to Gaia DR3 Synthetic Photometry catalogue \citep{gaiadr3syn} i.e. they are the average values for the period of observations between July 2014 and May 2017. Also included are the distances \citep{Bailer-Jones2021}, proper motions \citep{GaiaEDR3}, and sizes of the nebulae. The values of the Gaia Renormalized Unit Weight Error (RUWE), which can be used as an indicator for unresolved binary systems \citep{RUWE2024}, is on the order of unity for both stars.  However, with distances to both objects greater than 1\,kpc, such a RUWE value does not exclude scenarios of either close or long-period binary systems \citep{RUWE2024}. 

\begin{figure}
\centerline{
\resizebox*{0.5\columnwidth}{!}{\includegraphics[angle=0,viewport=15 30 750 750,clip]{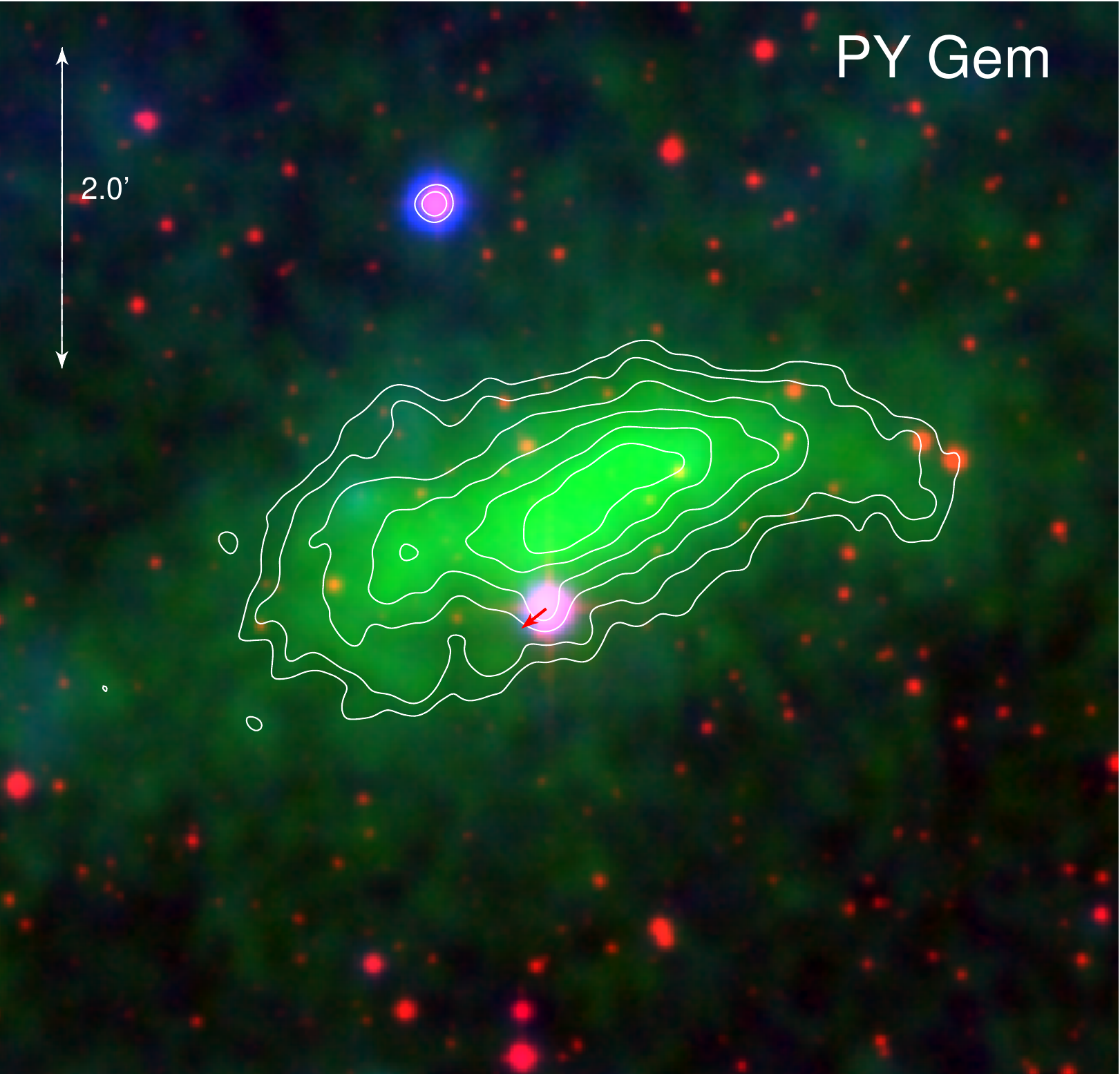}}
\resizebox*{0.5\columnwidth}{!}{\includegraphics[angle=0,viewport=15 24 750 750,clip]{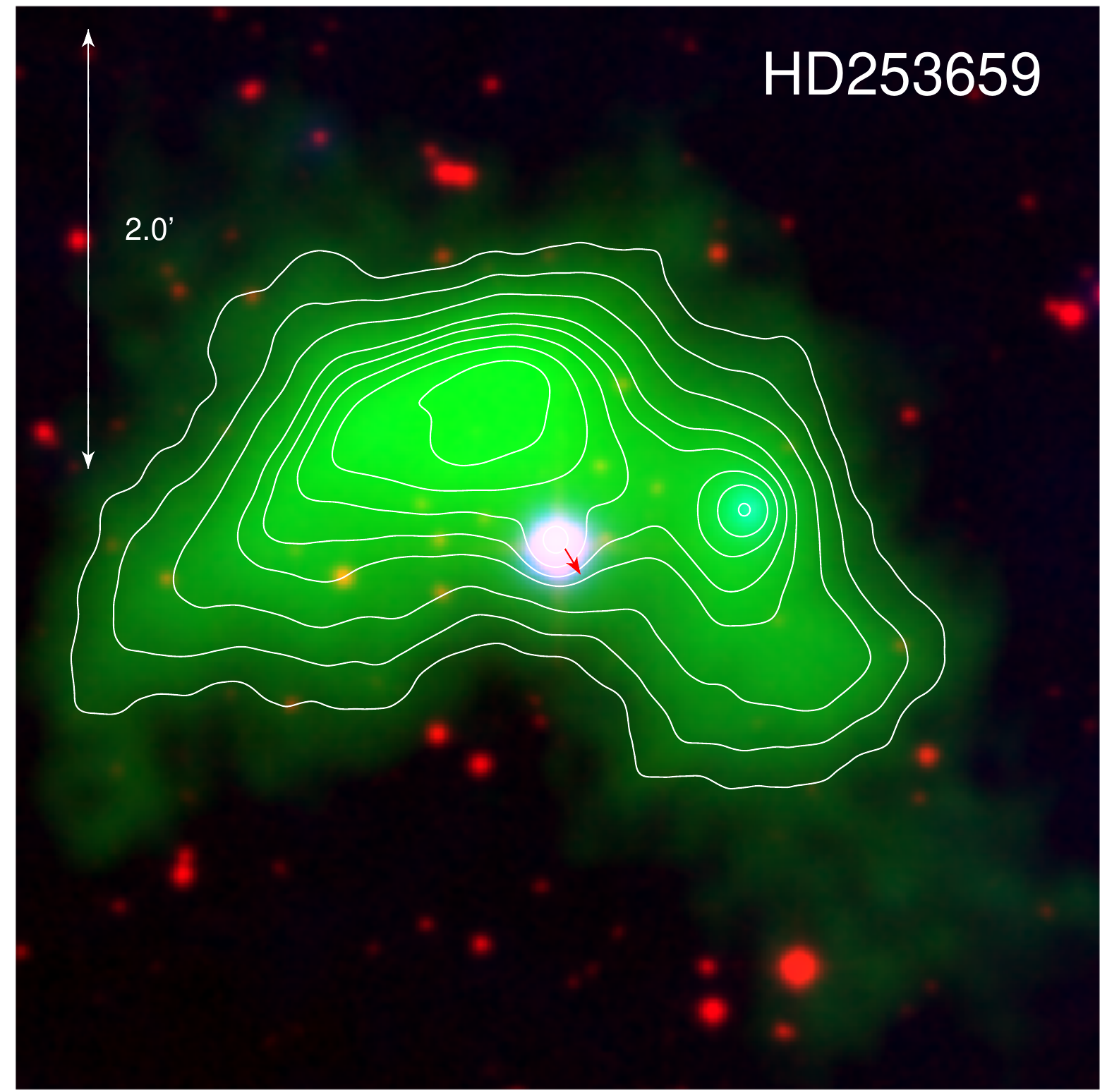}}}
\caption{RGB images with optical DSS-2-red (red channel), {\it WISE} $W4$ 22-$\mu$m (green channel), and WISE $W3$ 12-$\mu$m (blue channel) bands showing structures of the discovered nebulae. Contour lines represent the isophotes of the intensity distribution in the WISE $W4$ 22-$\mu$m band. Red arrows indicate the direction and amplitude of the central star proper motions for 10\,000~years. The vertical arrow on the left defines the image scale and has a length of 2 arcminutes, which is 10 times larger than the {\it WISE} $W4$ band angular resolution  (12$''$).}
\label{fig:nebulae}
\end{figure}

Figure~\ref{fig:nebulae} shows pseudocolour images of the nebulae constructed from optical and infrared data. The nebula of PY\,Gem is extended, and the star is located at its edge, while the one around \hdtwentyfive\ has an arc-like structure similar to a bow shock or a part of a circular nebula, probably indicating the heterogeneity of the ISM.  Although the nebula near PY\,Gem is not perfectly centered on the star, no other possible sources are detected in the region. Moreover, the physical size of the nebula, estimated assuming the distance to PY\,Gem, supports its connection with the star.

 With their classification as eruptive variables, \hdtwentyfive\ and PY\,Gem are ideal objects to investigate their activity over a long temporal baseline to search for possible indications for a connection between  disk formation processes and pulsational activity. Therefore, we have carried out a detailed photometric and spectroscopic study of these objects. The paper is organised as follows. In Section~\ref{sec:photometry} we describe the long-term and short-term photometric variability of the objects using ground-based and space photometry. We present our spectroscopic observations in Section~\ref{sec:spectra}. In  Section~\ref{sec:results}  we derive the stellar parameters, characterize the motion of \HDtwentyfive, in the colour-colour and colour-magnitude diagrams in terms of formation and dissipation of the disk,  and analyze the detected frequencies. In Section~\ref{sec:discussion} we discuss the outburst phases and disk-forming activity in relation to the observed nebulae, and we conclude our work in Section~\ref{sec:conclusions}.

   \begin{figure*}
   \centering
   \includegraphics[width=2.0\columnwidth,clip]{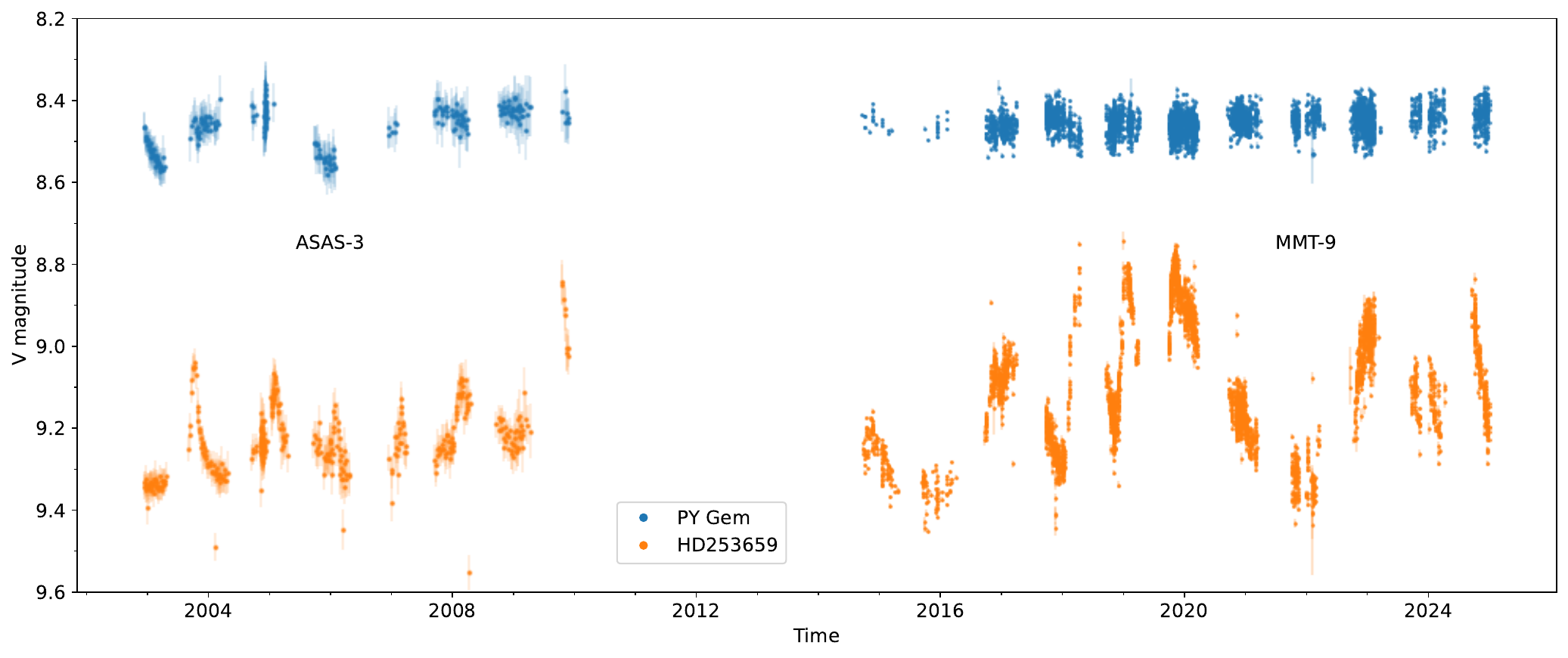}
   \caption{Light curves of PY\,Gem\ (blue dots) and \HDtwentyfive \ (orange dots) from the ASAS-3 and Mini-MegaTORTORA  sky survey archives. Formal photometric accuracy of individual measurements is represented with error bars, for most points it is comparable to the marker size.\label{mmt}}
   \end{figure*}

\section{Photometry}\label{sec:photometry}
\subsection{Optical counterpart}
\label{sect:optical-im}

We conducted a search for possible optical counterparts of the discovered IR nebulae. The region around \HDtwentyfive\ is covered by the publicly released data from the INT Photometric H$\alpha$ Survey \citep[IPHAS;][]{iphas_dr2}. This survey is imaging the Northern Milky Way in visible light ($r$, $i$ and H$\alpha$ filters) using the Isaac Newton Telescope (INT) in La Palma, Spain. We downloaded and stacked the H$\alpha$ images around this star from the IPHAS data archive\footnote{The IPHAS data archive is available online at \url{https://www.iphas.org}.}, but did not detect any signs of extended emission on the scales comparable to the extent of infrared nebulae.

As PY\,Gem is not covered by IPHAS images, we performed a set of dedicated observations 
with the Small Binocular Telescope \citep[BART/SBT;][]{BART2019} at Ond\v{r}ejov observatory  of the Astronomical Institute of the Czech Academy of Sciences. BART/SBT is a robotic telescope system of two 0.2-m wide-field optical telescopes equipped with a set of Sloan and nebular filters. We acquired 200 180-s exposure frames (10 hours in total) in H$\alpha$ over 4 nights in March 2025. We calibrated and stacked all the images, but also did not find any signs of extended H$\alpha$ emission around the star. 

   \begin{figure*}
   \centering
   \includegraphics[width=2.0\columnwidth,clip]{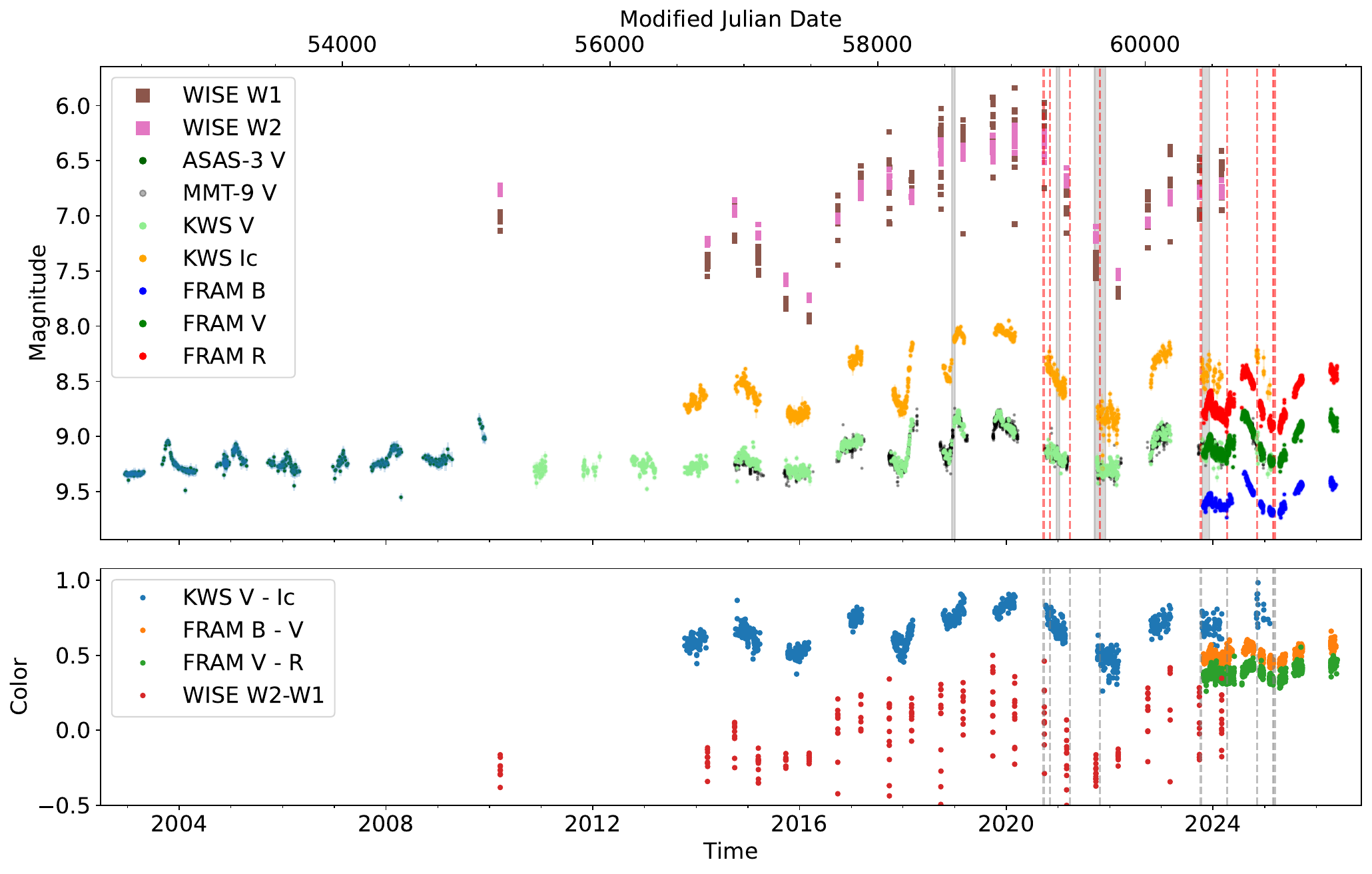}
   \caption{{\it Upper panel:} light curve of \HDtwentyfive\ over the last 20 years using the data from ASAS-3, KWS, and Mini-MegaTORTORA (MMT-9) sky surveys and our dedicated FRAM observations. {\it Lower panel:} variability of the colours of \HDtwentyfive. Vertical dashed lines mark the times of spectral observations, while gray bars indicate the intervals covered by \tess. \label{kws}}
   \end{figure*}
   \begin{figure*}
   \centering
   \includegraphics[width=2.0\columnwidth,clip]{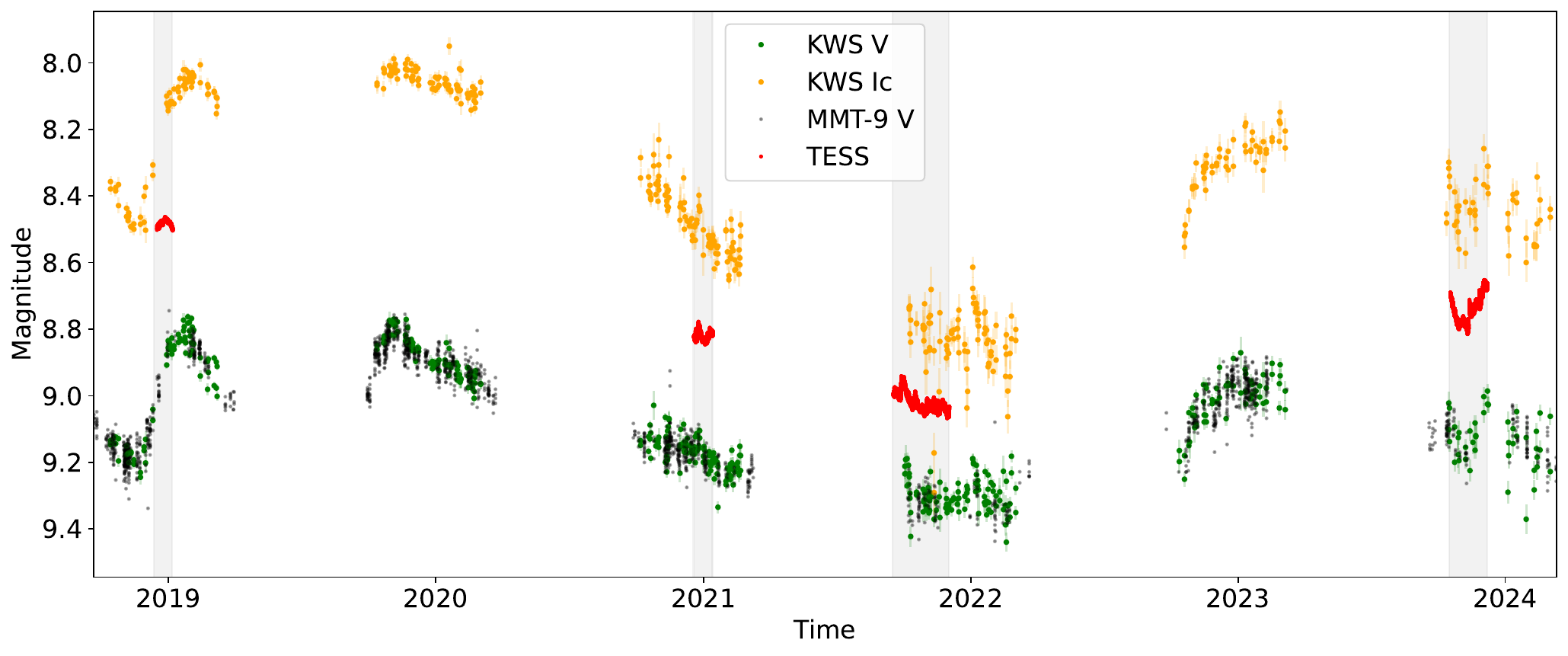}
   \caption{The light curves of \HDtwentyfive\ in the V and Ic bands of KWS, and in the V band of Mini-MegaTORTORA (MMT-9), around the time intervals covered by TESS (gray vertical bars, like in Figure~\ref{kws}). TESS light curves are presented in magnitudes directly derived from instrumental fluxes (see Section~\ref{sec:tess_short}), and arbitrarily shifted vertically for better comparison. 
   \label{kws_tess}}
   \end{figure*}
\begin{table}\centering
\caption{Summary of photometric data used in the study. MMT-9 is Mini-MegaTORTORA, FRAM is FRAM-ORM. N is the number of measurements. \label{tab:photometry}}
\resizebox{\columnwidth}{!}{\begin{tabular}{lcc cc}
\hline
\multirow{2}{*}{Star}  & \multirow{2}{*}{Telescope} & \multirow{2}{*}{Observing Interval}  & \multirow{2}{*}{Filter} & \multirow{2}{*}{N} \\
\\
\hline
\multirow{2}{*}{PY\,Gem}  &   ASAS\,3  & \multicolumn{1}{l}{2002-12-14 - 2009-11-30} & $V$ &  262 \\   
                          &   MMT-9    & \multicolumn{1}{l}{2014-09-23 - 2024-12-31} & Clear &  4012 \\ 
\\
\multirow{5}{*}{\HDtwentyfive}      &   ASAS\,3   & \multicolumn{1}{l}{2002-12-13 - 2009-11-30}  & $V$          &  395 \\
                                    & {\it WISE}   &  \multicolumn{1}{l}{2010-03-17 - 2024-03-02} & W1, W2 & 250 \\
                                    &   KWS       & \multicolumn{1}{l}{2010-12-04 - 2025-02-14}   &$V$, $Ic$     & 2038  \\
                                    &   MMT-9     & \multicolumn{1}{l}{2014-09-30 - 2024-12-31}  & Clear  & 3348  \\
                                    &   FRAM  & \multicolumn{1}{l}{2023-10-26 - 2026-05-27}   &$B$, $V$, $R$ & 3772  \\ 
\hline
\end{tabular}}
\end{table}

   \begin{figure*}
   \centering
   \includegraphics[width=18cm]{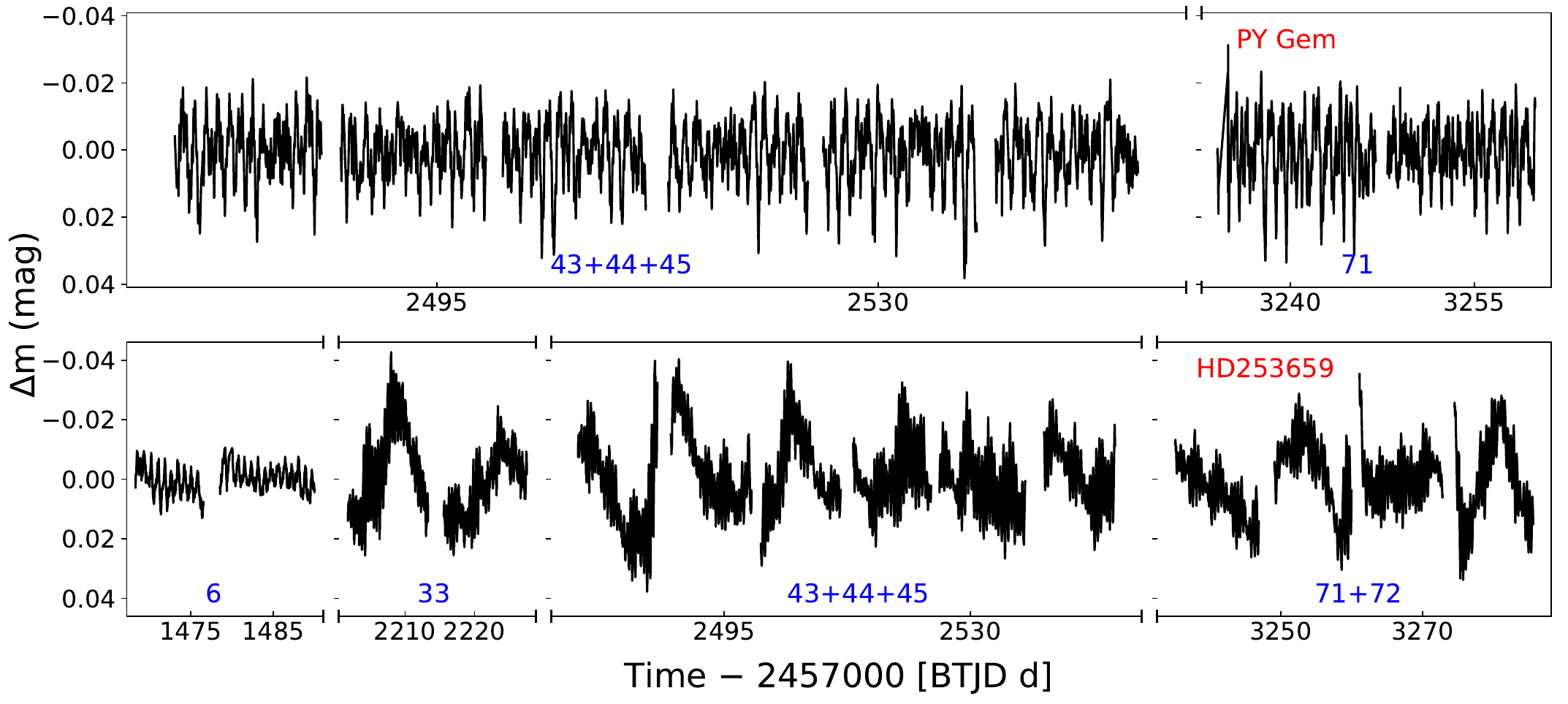}
   \caption{\tess\ light curves of PY\,Gem ({\it upper panel}) and \HDtwentyfive\  ({\it lower panel}) in different sectors (blue labels).  \label{tess}}
   \end{figure*}

\subsection{Long-term behaviour}

In order to assess the long-term variability of the stars, we acquired their photometry from the data archive of Mini-MegaTORTORA \citep{MMT2017}, which is a nine-channel wide-field optical monitoring system with high temporal resolution, operating since mid-2014 and located at the Special Astrophysical Observatory, Nizhny Arkhyz, Russia. It performs a systematic monitoring of the northern sky, covering every point of it on average several times per night in white light. The photometric measurements are then re-calibrated to Johnson $V$ band by a dedicated pipeline that, apart from standard calibration steps, determines the effective photometric system of every frame, and then employs this information to derive the $(B-V)$ colours of every star \citep{Karpov2018}. The resulting measurements are published online on the dedicated portal\footnote{\url{http://survey.favor2.info/}}. The data for the stars have been extracted from the Mini-MegaTORTORA archive, and passed through quality cuts in order to filter out the points corresponding to bad weather intervals and images where photometric calibration was too noisy. This resulted in more than 3300 points with good quality $V$ magnitudes for every star, spanning more than 10 years since mid-2014 (Table~\ref{tab:photometry}). 
Additional photometric data covering the years 2003-2010 were also acquired from the archive of the northern observational station of the All Sky Automated Survey  \citep[ASAS-3,][]{ASAS3Pojmanski1997}. 

Figure~\ref{mmt} shows the light curves constructed from the Mini-MegaTORTORA and ASAS-3 data. \hdtwentyfive\ shows long-term irregular variability with an amplitude of 0.4~mag. Such a photometric behaviour agrees with the classification of \hdtwentyfive\ as a $\gamma$\,Cas variable. In contrast, the light curve of PY\,Gem, also classified as $\gamma$\,Cas variable, only shows noticeable variability about 20 years ago between 2003 and 2008. In the last $\sim 10$\,yr, PY\,Gem displays only low-amplitude ($< 0.05$\,mag) variability.

With the aim of studying the colour behaviour of \HDtwentyfive, we initiated multi-colour photometric observations at the end of October 2023 on FRAM-ORM, which is a 10-inch Meade f/6.3 Schmidt-Cassegrain telescope. The FRAM-ORM telescope is installed in the Roque de Los Muchachos Observatory, La Palma, and equipped by the Moravian Instruments G2-1600 CCD. Data were acquired in Johnson-Cousins  $B$, $V$, and $R$ filters with 30, 20 and 20 seconds exposures, respectively, to avoid saturation due to the brightness of the star, and then automatically processed by a dedicated Python pipeline based on the {\sc STDPipe} package \citep{stdpipe}, which includes bias and dark current subtraction, flat-fielding, cosmic ray removal, astrometric calibration, aperture (with a 5-pixel radius) photometry, and photometric calibration using the catalogue of synthetic photometry based on Gaia DR3 low-resolution XP spectra \citep{gaiadr3syn}. The upper panel of Figure~\ref{fram} presents the light curves in the $B$, $V$ and $R$ bands, while the lower panel shows the temporal variations of the colours $B-V$ and $V-R$. 

To complement these data, we used infrared photometry for \HDtwentyfive\ from the AllWISE \citep{allwise} and NEOWISE Reactivation Mission \citep{neowise} single exposure catalogues based on {\it WISE} \citep{wise} observations in the W1 and W2 bands. We also found continuous observations of \HDtwentyfive\ in the $V$ and $Ic$ bands in the Kamogata/Kiso/Kyoto Wide-field Survey (KWS, \citet{Maehara2014}).  Figure~\ref{kws} presents the combined light curve of \HDtwentyfive\  since 2003, along with the evolution of several colours based on the data outlined above.

\begin{table}
\caption{Summary of \tess\ observations. 
Listed are the observing sector (S), corresponding time interval, and the cadence of observations. \label{tab:tess_obs}}
\resizebox{\columnwidth}{!}{
\begin{tabular}{lccc}
\hline 
\multirow{2}{*}{Object} & \multirow{2}{*}{S} & \multirow{2}{*}{Observing Interval} & Cadence   \\
& & & [s] \\
\hline
\multirow{4}{*}{PY\,Gem}  & 43 & 2021-09-16 - 2021-10-12 & 600 \\ 
         & 44 & 2021-10-12 - 2021-11-05 & 600 \\  
         & 45 & 2021-11-06 - 2021-12-02 & 600 \\
         & 71 & 2023-10-16 - 2023-11-10 & 200\\
         &    &                         &   \\
\multirow{7}{*}{\HDtwentyfive} &  6 & 2018-12-11 - 2019-01-07 & 1800 \\
              & 33 & 2020-12-18 - 2021-01-13 & 600 \\
              & 43 & 2021-09-16 - 2021-10-12 & 600 \\
              & 44 & 2021-10-12 - 2021-11-05 & 600 \\
              & 45 & 2021-11-06 - 2021-12-02 & 600 \\
              & 71 & 2023-10-16 - 2021-11-11 & 200 \\
              & 72 & 2023-11-11 - 2021-12-07 & 200 \\
\hline
\end{tabular}
}
\end{table}

\subsection{Short-term variability}
\label{sec:tess_short}

PY\,Gem and \HDtwentyfive\ have been monitored with \tess\ \citep{tess} at various cadences (Table\,\ref{tab:tess_obs}) and in different sectors simultaneous with the ground-based observations as shown in Figure~\ref{kws} and \ref{kws_tess} for \HDtwentyfive. We made use of the \textsc{TESScut} service \citep{2019ascl.soft05007B} to extract cutout grids from the TESS full-frame images (FFIs). To optimize modelling of the background signal, we set the cutout size to $20\times20$ pixels. We then proceeded to extract the stellar flux and correct against the systematics, using the build-in methods from the dedicated Python package Lightkurve \citep{2018ascl.soft12013L}. For measuring the target flux, we first defined an aperture of pixels with brightness exceeding 4~$\sigma$ that of the median value of the grid. Accordingly, a mask of pixels with brightness below the $10^{-4}~\sigma$ level was used for measuring the background signal. The extracted background vectors were then subject of dimensionality reduction using three principal components, and the best-fit noise models were subtracted from the flux of the two science objects.  Figure~\ref{kws_tess} shows a zoomed-in view of the light curves in Figure~\ref{kws}, including these TESS light curves, with fluxes converted to magnitudes with arbitrarily chosen zero point for easier visual comparison with other datasets. We can see that the TESS light curve follows a similar long term trend as the ground-based photometry.
To facilitate the extraction of periodic signals, we correct for these long-term trends by additionally fitting and subtracting second-order polynomials from the fluxes in every sector of the data. The resulting light curves of our objects, in magnitudes relative to the fit baseline, are displayed in Figure~\ref{tess} and are used for further analysis. 

The \tess\ data of PY\,Gem show high-frequency variability that is similar across all sectors. In contrast, \HDtwentyfive\ displays rather erratic variations. The most remarkable is the abrupt change in the type of variability when comparing the \tess\ data from sector 6, where the light curve has a significantly lower amplitude\footnote{This finding is also confirmed by inspecting the instrumentally corrected products (so called PDCSAP flux) produced by the Science Processing Operations Center (SPOC) pipeline, which are available at the Barbara A. Mikulski Archive for Space Telescopes.} and the behaviour resembles a bit that of PY\,Gem, with those of all other sectors, where the high-frequency variability appears to be superimposed on longer, more chaotic signals.

We used \tess-Localize \citep{2023AJ....165..141H} to verify the source of the detected variability. 
This software uses the observed frequencies as input parameters and calculates the best-fit locations for the signal sources within the provided Target Pixel File (TPF). Our analysis confirms that HD253659 is the most likely Gaia source with $G \leq 18$\,mag that displays the observed variability within the $25\times25$-pixel TPF, with a relative likelihood greater than 99$\%$.

\begin{figure*} 
\centerline{
\resizebox*{2.0\columnwidth}{!}{\includegraphics[angle=0]{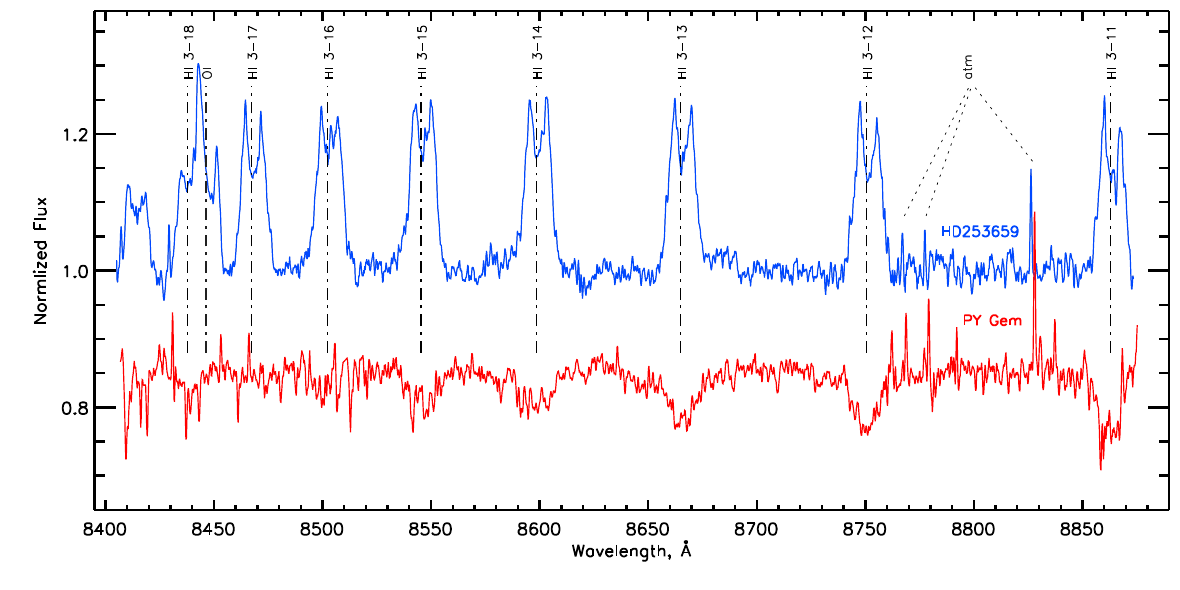}}
}
\vspace{-0.5cm}
\caption{Near-IR spectra of PY\,Gem and \HDtwentyfive\ obtained with the Perek 2-m telescope in October 2023 and April 2021, respectively. \label{fig:gaiaspectrum}}
\end{figure*}   
\begin{figure*}
\centerline{
\resizebox*{\textwidth}{!}{\includegraphics[angle=0]{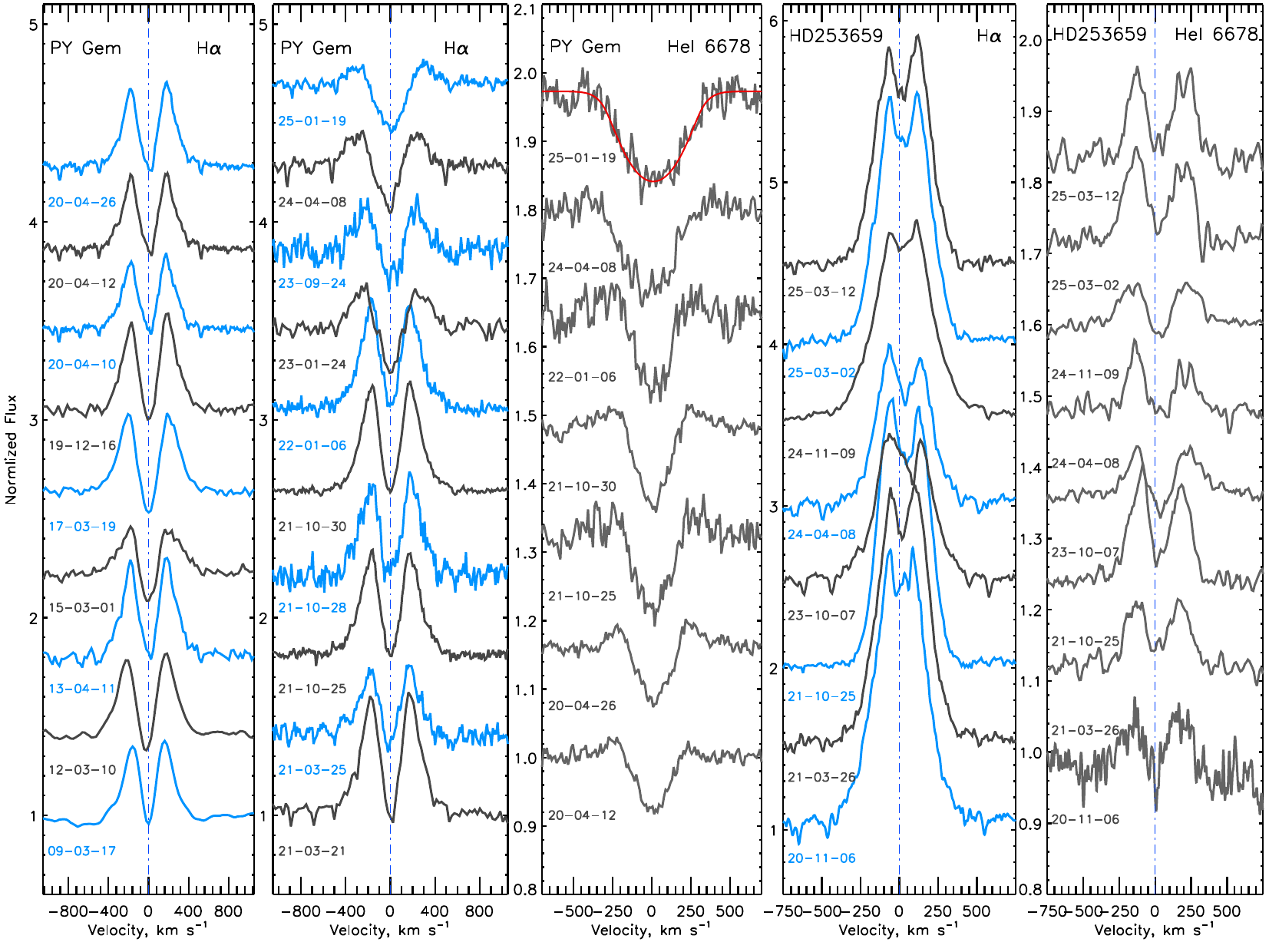}}
}
\vspace{-0.1cm}
\caption{Variability of line profiles in the spectra of PY\,Gem ({\it panels 1-3 from the left}) 
and \HDtwentyfive\ ({\it panels 4-5}). 
The red line 
in the third panel shows the result of the line-broadening analysis with the {\sc iacob-broad} routine. Dates of observations are given in the format (year-2000)-month-day.  
\label{fig:HaHD}}
\end{figure*}   

\section{Spectral observations}\label{sec:spectra}

With the aim of investigating the H$\alpha$ line profile variability, we initiated  
spectral monitoring of the stars at the Perek 2-m telescope, operated by the Astronomical Institute of the Czech Academy of Sciences. Since 2020 we obtained 11 medium-resolution spectra ($R \simeq 13{,}000$) covering the H$\alpha$ line for PY\,Gem  and 8 spectra for  \HDtwentyfive.  Moreover, for both stars we acquired medium-resolution spectra in the near-IR range (8400-8880~\AA) covering higher order lines of the Paschen series of hydrogen using the same coude spectrograph (Figure~\ref{fig:gaiaspectrum}) and one high-resolution spectrum  ($R \simeq 50{,}000$) covering the whole optical range for PY\,Gem using the Ond\v{r}ejov Echelle Spectrograph (OES; \citealp{OES2004,OES2020}).  The spectra from both spectrographs were reduced using a dedicated IDL-based pipeline in the same way as it was done in \citet{Maryeva2022, Maryeva2023}. After primary reduction, all spectra were normalized and corrected for barycentric velocity. 

We complemented our data set with the spectra of \hdtwentyfive\ and PY\,Gem retrieved from the BeSS database \citep{Neiner2011, Neiner2018}. The dates of all these  spectral observations and the durations of exposures are summarized in Table~\ref{tab:log11}. 

Figure~\ref{fig:HaHD} presents the variability of the H$\alpha$ and \HeI$\lambda6678$ profiles in the spectra of PY\,Gem and \HDtwentyfive, while Tables~\ref{tab:PYGem} and~\ref{tab:HD25} list the measured parameters of the H$\alpha$ lines (intensities of blue and red peaks, radial velocity). 
The distance between the red and blue peaks of the double peaked H$\alpha$ line shows that PY\,Gem has a significantly higher disk inclination to the line of sight compared to \hdtwentyfive, assuming the rough dependence of $v\sin{i}$ on the inter-peak distance from \citet{Hanuschik1989} and that the stars have similar values of $v$ due to belonging to contiguous spectral types and comparable widths of their H$\alpha$ line wings ($\approx800$\kms). 

The positions of the red and blue peaks in the spectra of both stars vary from one observation to another (Tables~\ref{tab:PYGem} and~\ref{tab:HD25}). However, in the collected spectral data we do not detect any systematic radial-velocity variations that could be interpreted as evidence of binarity.

The central panel of Figure~\ref{fig:HaHD} shows that, in the case of PY\,Gem, the \HeI$\lambda6678$ profile is a superposition of photospheric absorption and double-peaked emission originating from the disk. Only in the spectrum obtained in January 2025 the disk contribution disappeared, and we observe pure photospheric absorption in \HeI$\lambda6678$. We used this profile to measure the projected rotational velocity of PY\,Gem. For this, we applied the {\sc iacob-broad} {\sc idl} procedure developed by S.~Sim{\'o}n-D{\'\i}az \citep{SimonDiaz2014iacob-broad}, which simultaneously estimates both $v\sin{i}$ and the macroturbulent velocity $v_m$. The {\sc iacob-broad} routine is based on a combined Fourier Transform (FT) and goodness-of-fit (GOF) approach, where an intrinsic line profile is convolved with various broadening functions and compared to the observed profile using a $\chi^2$ minimization method \citep{SimonDiaz2014iacob-broad}. From this analysis, we derived a projected rotational velocity of $v~\sin{i}=303\pm15~$\kms \ and a macroturbulent velocity of $v_m=154\pm 15~$\kms. The value of $v~\sin{i}$ that we obtained is in good agreement with rotational velocities typical for Be stars \citep{Zorec2016Be_rotation}. We note, however, that the value for the macroturbulent velocity must be taken with caution and should be considered as an upper limit. Because the rotation is dominating the broadening of the profile, a possible contribution in form of a Gaussian profile (as is assumed for the macroturbulent velocity) will not be noticeable in the profile until it reaches a significant value.

In the spectrum of \hdtwentyfive\ all hydrogen lines in the optical and near-IR ranges are in emission and have double-peaked profiles. The oxygen \OI$\lambda8446.3$ line is clearly visible in the spectrum -- it also has a double-peaked profile and overlaps with the hydrogen H\,{\sevensize I}~\,3-18 line. In addition to the hydrogen and helium lines in the optical range, we also found a broad double-peaked line of Na\,{\sevensize I}\,$\lambda$5688.2. In the spectrum of \hdtwentyfive\ there are no pure photospheric absorption lines, which can be used to estimate its projected rotational velocity. 

It is important to note that we did not detect any signatures of nebular lines in the spectra of either of the two stars. This is consistent with the results presented in Sect.\,\ref{sect:optical-im}, where no evidence of extended H$\alpha$ emission was found.

\section{Results}\label{sec:results}
\subsection{Stellar parameters}\label{sect:stelparam}

In the spectra of both stars we detect significant contribution from the circumstellar disk to the observed line profiles, which prevents a reliable determination of the stellar parameters through spectral line fitting. Therefore, we derived the stellar parameters using spectral energy distribution (SED) fitting.  This approach is widely applied to Be and evolved massive stars  \citep{Arcos2018, Kourniotis2018}. 

To construct the SEDs, we assembled multi-colour photometric measurements spanning from the $U$ band to the infrared using data from various photometric catalogues\footnote{The photometric measurements are taken from:  WISE \citep{allWiseCutri}; Two Micron All Sky Survey (2MASS) \citep{Cutri2MASS};  {\it Gaia} DR3 \citep{gaiadr3syn}; the Panoramic Survey Telescope and Rapid Response System (Pan-STARRS) DR1 \citep{PanSTARRS2016}, \citet{Mermilliod1987}.} available through VizieR \citep{vizier}. We note that these measurements were obtained at different epochs and that both targets are photometrically variable. However, no simultaneous multi-wavelength observations are available.  Therefore, for both stars, we compiled photometric measurements obtained at different epochs but corresponding to the same quiescent state. To represent the quiescent state of \HDtwentyfive, we used the $U$- and $B$-band measurements from 1973, 2MASS photometry obtained shortly before 2000, Pan-STARRS photometry (except the $g$ band) obtained between 2010 and 2014, and the average \textit{WISE} W1 and W2 fluxes from March 2016, corresponding to the global minimum of the \textit{WISE} light curve. To represent the quiescent state of PY\,Gem, we excluded the 2MASS data from the analysis because, according to the ASAS-3 light curve, the star underwent a decrease in brightness in the early 2000s.
%
%
 The photometric uncertainties adopted in the SED fitting were chosen to account not only for the quoted catalogue errors but also for the expected scatter introduced by the intrinsic variability of the stars. Under these assumptions, the assembled photometry provides a reasonable representation of the average SED of each target. 
We used the models from the grid of stellar continuum spectra provided by \citet{1992IAUS..149..225K} and interstellar reddening with the extinction curve of \citet{1989ApJ...345..245C} with a value of $R_{V} = 3.1$ to represent these SEDs. The distances used are listed in Table~\ref{tab:stars}. 

\begin{figure}
  \centering
  \includegraphics[width=\columnwidth]{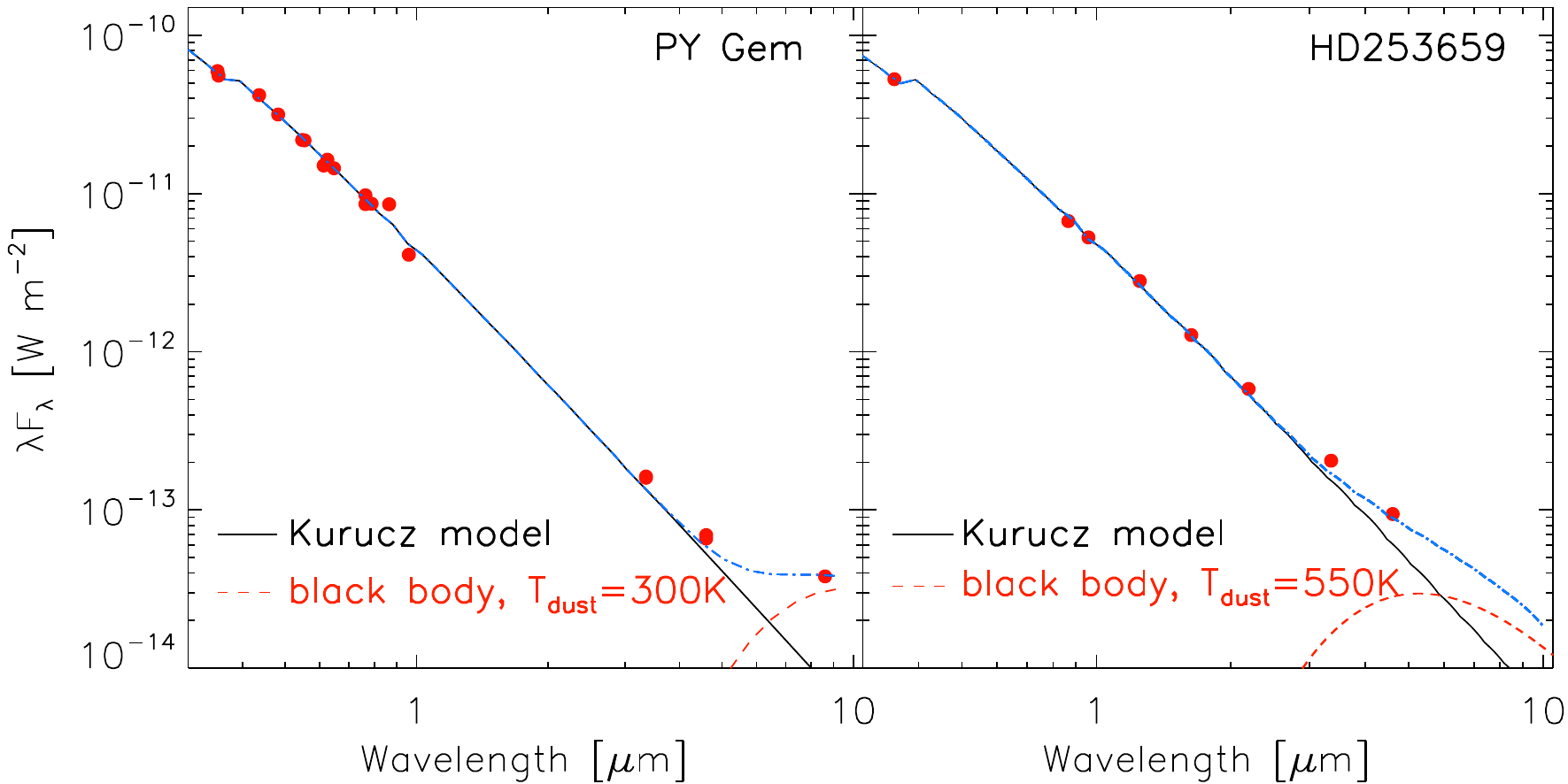}
  \caption{Stellar continuum fits (solid lines) to the observed, dereddened photometric data (red points) of PY\,Gem (left) and HD\,253659 (right). The best-fitting parameters are listed in Table~\ref{tab:parameters-SED}.  Warm dust (red dashed line) contributes to the SEDs of both stars.} 
  \label{fig:SED} 
\end{figure}

We followed the recipe of \citet{MaizApellaniz2004} for obtaining the best fitting models and the reddening. The sets of stellar parameters and the interstellar extinction values obtained from the fit are given in Table~\ref{tab:parameters-SED}. The dereddened data along with the best-fitting models are shown in Figure~\ref{fig:SED}.  The SED of PY\,Gem shows that dust emission becomes dominant at wavelengths closer to 10\,$\mu$m, corresponding to temperature of  $T_{\rm d} \sim 300$\,K. For HD\,253659 the infrared excess emission is visible at wavelengths greater than 3\,$\mu$m and can be represented with a Planck function with a temperature of $T_{\rm d} \sim 550$\,K. We propose this excess emission to be due to warm dust, because in quiescence the star should not have a disk emitting free-free emission, but we cannot exclude that the mid-infrared emission might emerge from a disk remnant. Unfortunately, no WISE 3 and 4 band observations exist during quiescence, which could help discriminating between the two scenarios.

\begin{figure}
  \centering
  \includegraphics[width=\columnwidth]{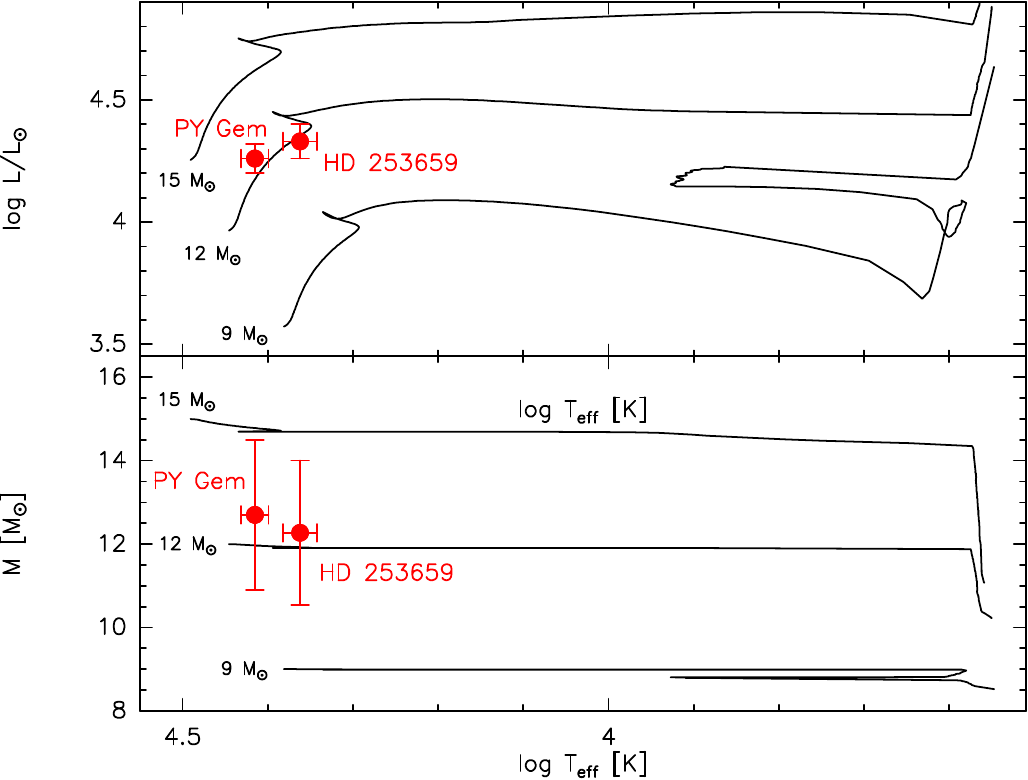}
  \caption{Positions of the stars in the HR diagram (top). A comparison of the obtained masses of the objects (dots) with evolutionary predictions (solid lines) is shown in the bottom panel. Evolutionary tracks are from \citet{Ekstrom} for rotating stars at solar metallicity.}
  \label{fig:evol} 
\end{figure}

In the top panel of Figure~\ref{fig:evol}, we show the location of the objects in the Hertzsprung-Russell (HR) diagram together with stellar evolutionary tracks from \citet{Ekstrom} for rotating stars with initial masses in the range $9-15$\,M$_{\sun}$ at solar metallicity.  
Although the derived $\log g$ is very sensitive, on the one hand, to the grid step of the models and, on the other hand, to the height of the Balmer jump and thus to the $U$-band flux, we find reasonably good agreement between the mass estimated from the SED and the one from evolutionary tracks (bottom panel of Figure~\ref{fig:evol}). 

\begin{table*}\centering
\caption{Derived sets of stellar parameters. \label{tab:parameters-SED} }
\begin{tabular}{lcc ccc cc }
\hline
Star & $T_{\rm eff}$, [K] &  $\log T_{\rm eff}$, [K] & $\log g$, [cgs] & $R$,  [$\Rsun$]     &$\log L/\Lsun$ & $M$, [$\Msun$]      & $A_V$, [mag]  \\
\hline
  PY\,Gem             &      $26300\pm1200$              &  $4.42\pm0.02$  &  $3.9\pm 0.1$ &   $~\,6.62\pm 0.24$     & $4.26\pm0.06$ & $12.7\pm1.8$   & $1.05\pm0.02$  \\
  \HDtwentyfive       &   $23000\pm1000$                 &  $4.36\pm0.02$ & $3.6\pm 0.1$ &       $9.19\pm0.32$  &  $4.33\pm0.07$ & $12.3\pm1.7$ & $1.92\pm0.02$  \\
\hline
\end{tabular}
\end{table*}

\subsection{ Major outburst of  \HDtwentyfive\ and its disk evolution}
\label{sect:outburst}

Be stars may demonstrate major outbursts  associated with the complicated process of circumstellar disk formation. Major outbursts are observed as brightenings in broadband photometry (usually up to 0.5-0.6 mag) and may have different durations \citep{Klaus2018Bevariability}: from a few tens of days \citep[for example BD+61\,39;][]{Labadie-Bartz2018outbursts} up to a few years  \citep[for example 2dFS0231;][]{Michalis2014}, $\omega$\,CMa \citep{Ghoreyshi2018}, KIC\,9715425 \citep{Gao2025Bemajoroutburst}, EPIC\,202060631 \citep{GaoYuan2025}). Numerical simulations have shown that Be stars should turn redder at the beginning of the major outburst phase due to the formation of a viscous decretion disk \citep[see, e.g.,][]{Haubois2012}. 

The light curve of \HDtwentyfive\ (Figure~\ref{kws}) clearly shows that the star experienced a series of minor outbursts between 2003 and 2010 with brightening in $V$ band of $0.2-0.3$\,mag during each event. 
These minor outbursts, when compared to theoretical predictions of \citet{Haubois2012}, could be interpreted with (quasi-)periodic mass ejections with a (quasi-)period of $\sim 1.16$\,yr and duty cycles (defined as the fraction of time spent actively ejecting mass) in the range $\sim 10-20$\,\%. The clear brightening in the $V$ magnitude during these events speaks in favour of a more pole-on orientation ($i \la 30\degr$) of the disk.

The broad bump with a total brightness increase of $>0.4$\,mag observed between 2016 and 2022 may be interpreted as a major outburst.  After 2016 we see two maxima in the light curve and a plateau from 2019 till 2021. The jagged plateau resembles those of $\omega$\,CMa during its major outbursts in 1982-1987 or 2009-2011 \citep{Ghoreyshi2018}. But in contrast to the cyclic outbursts of $\omega$\,CMa that show an instantaneous rise in magnitude (occasionally followed by a short series of brightness increase and dimming episodes indicating alternating build up and dissipation phases) followed by a plateau phase of nearly constant brightness and a steep drop in magnitude that marks the onset of the subsequent quiescent (i.e. disk dissipation) phase, the event monitored in \HDtwentyfive\ is less steep at both the beginning and the end of the event and shows deeper minima during the plateau phase, reaching even the pre-outburst level. 
 This could indicate that the new disk that is formed has a lower viscosity than the previous minor outbursts \citet{Haubois2012}. During the plateau phase, the brightness shows also deeper minima than those reported for $\omega$\,CMa, reaching even the pre-outburst level, indicating that the entire disk-formation process is non-continuous with intermediate phases of dissipation. In addition, the stage of quiescence following the outburst event is significantly shorter (less than two years) before the next minor outburst sets in (during 2023) compared to what was reported for $\omega$\,CMa. 

According to multi-band photometry, \HDtwentyfive\ becomes redder during the major outburst, which is consistent with theoretical expectations and was also seen during the cyclic outbursts of $\omega$\,CMa \citep{Ghoreyshi2018}. The major outburst is observed at all wavelengths and is more pronounced in the IR range,  in agreement with the fact that longer wavelengths trace larger radii (and hence also a larger emitting area) within the disk, especially if it is seen close to pole-on \citep{Haubois2012}. The brightening in the {\it WISE W1, W2} bands is about 1.5~mag. We do not see any delay of the IR maximum of brightness in comparison with the one in the optical range. However, the decline begins in the $V$ band, while the infrared bands respond with a slight delay (see Figure~\ref{kws_tess} around 2020). This delay produces an inversion in the relationship between $V$ and $V-I_{\rm C}$. It can be explained by the clearing of the innermost ($r \la 2$\,R$_{*}$) and densest disk region, in which the excess in $V$-band flux is generated. 

\begin{figure*}
  \centering
  \includegraphics[width=2\columnwidth]{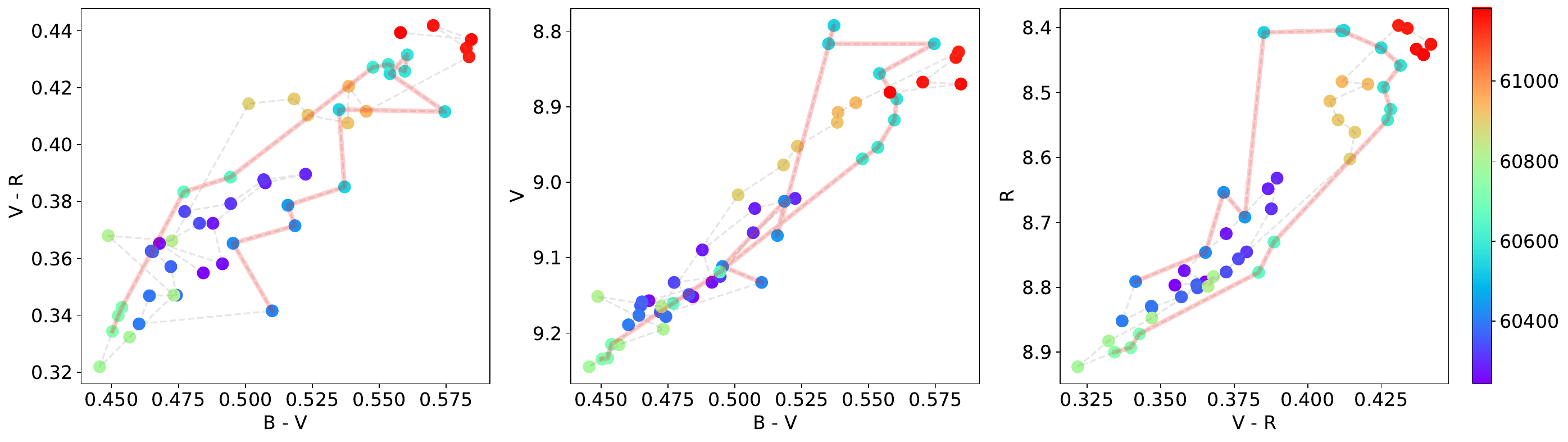}
  \caption{Colour-colour (left) and colour-magnitude (middle and right) diagrams of \HDtwentyfive\  based on FRAM multi-band photometry during the observing period 2022-2026. The colour bar shows Julian date of observations. The line in light red marks  the outburst detected between April 2024 and April 2025.  } 
  \label{fig:colorcolor} 
\end{figure*}

\begin{figure}
  \centering
  \includegraphics[width=1\columnwidth]{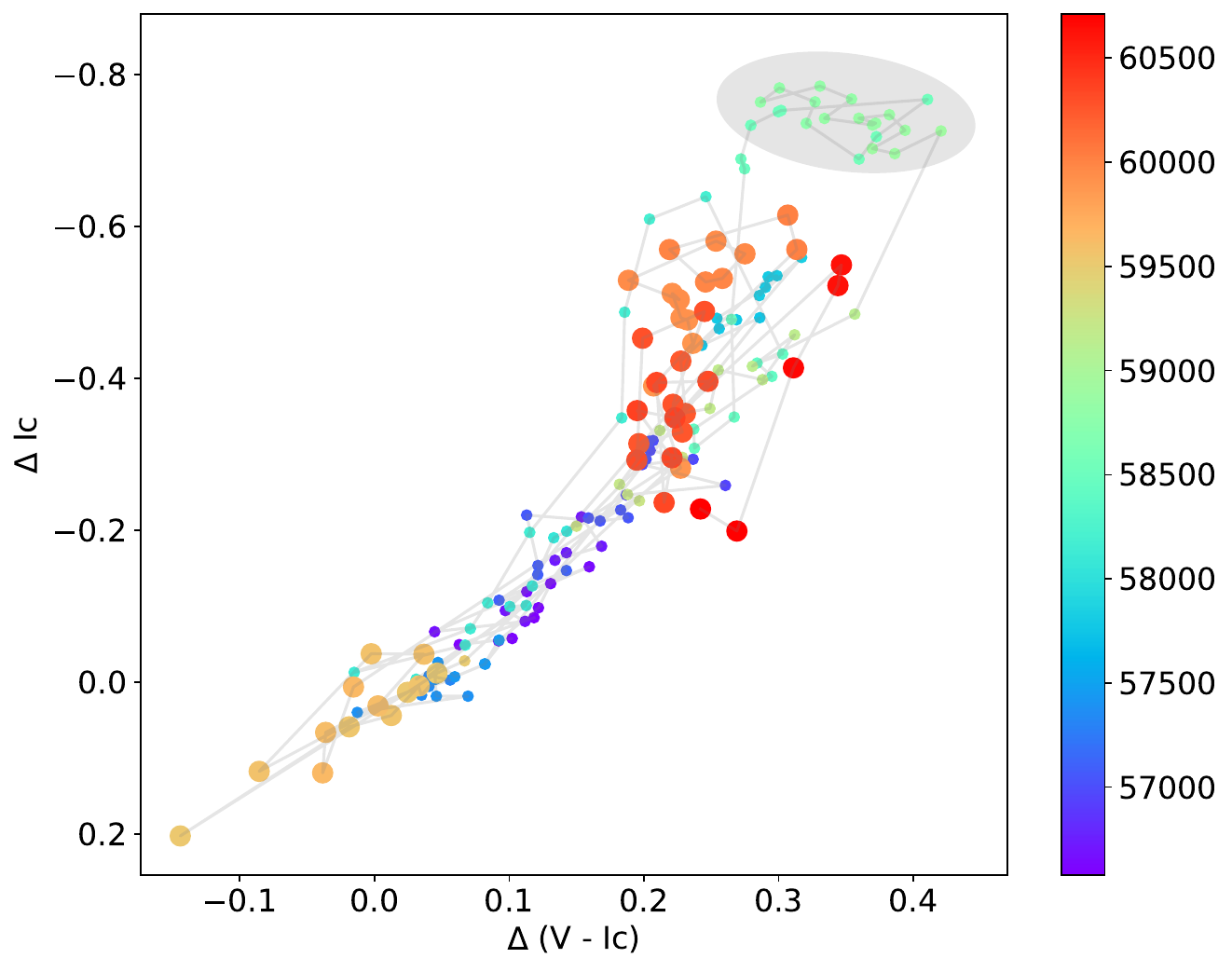}
  \caption{Colour--magnitude diagram  of \HDtwentyfive\  based on all KWS data (2014-2025). The small-sized symbols correspond to observations taken before October 2021. The zero points of magnitude and colour are the mean values over the light curve minimum around Jan 2022, which we consider to be most similar to the quiescent phase over the interval covered by KWS observations.} 
  \label{fig:colorcolorJapan} 
\end{figure}

To correlate the observed photometric variability with disk formation and dissipation activities, we follow the temporal evolution of \HDtwentyfive\ in colour-colour and colour-magnitude diagrams.  Figure~\ref{fig:colorcolor} shows the evolution at optical wavelengths based on the FRAM multi-band photometry collected between 2022 and 2026, which is the subsequent phase after the major outburst. During this period, \HDtwentyfive\ moved across these diagrams with an increasing amplitude of variability, tracing three loop-like patterns and eventually returning to the upper-right region of the diagrams. The overall behaviour of \HDtwentyfive\ is qualitatively consistent with the predictions of \citet{Haubois2012} for the build-up and subsequent dissipation of a circumstellar viscous decretion disk. In this framework, the upward motion in the colour-magnitude diagram corresponds to disk growth, while the downward motion reflects disk dissipation. The observed hysteresis indicates that the disk properties are different in the formation and dissipation phases. A similar colour-magnitude behaviour was shown by  $\delta$\,Sco \citep{Jones2013}, which is explained by variation of the density of the inner and outer part of the disk, with density decreasing rapidly with increasing radius \citep{Suffak2020}. Interestingly, the object appears dimmer for some time beyond the end of the outburst (April 2025). This could indicate that mass accretion back onto the stellar surface does not occur exclusively within the disk plane but rather via some funnel flows, leading to temporary circumstellar extinction.

We also inspected the colour-magnitude behaviour including the $I_{\rm C}$-band. Figure~\ref{fig:colorcolorJapan} presents the colour-magnitude diagram constructed from the full KWS photometric dataset spanning approximately 12 years and including the whole major outburst. Comparison with theoretical predictions by \citet{Figueiredo2025} confirms the repeated episodes of disk build-up and dissipation in a system viewed nearly pole-on. The narrowness of the loop suggests a relatively low base density at the inner edge of the disk. The points located in the upper-right corner (marked by the grey ellipse) correspond to the middle of the major outburst ($\approx$2020), during which an inversion of the colour--magnitude relation was observed: the $(V-I_{\rm C})$ colour increased while the object simultaneously dimmed in $V$. As mentioned previously, this can be explained by the clearing of the innermost disk region. The last group of dots forms broad loops, which corresponds to the phase after major outburst. Broadening of the loops is evidence of increasing of base density \citep{Figueiredo2025} and its significant fluctuations over short time scales, in agreement with the variations seen at optical wavelengths (Figure~\ref{fig:colorcolor}).

Our regular spectroscopic monitoring of \HDtwentyfive\ began in 2020, when the star was already recovering from the major outburst event (i.e., after the plateau phase). There is no clear correlation between the equivalent width (EW) measurements of the H$\alpha$ lines (last column of Table~\ref{tab:HD25}) and the brightness of the star. For similar stellar magnitudes, the EWs display a variety of values depending on whether they were measured during the recovery from the major outburst or during the post-outburst phase of less pronounced activity. This may indicate a delay in the H$\alpha$ response, similar to what was observed in the Be system EPIC~202060631 \citep{GaoYuan2025}. The intensities of the red and blue peaks of H$\alpha$ are similar ($Ib/Ir \sim 1$), indicating that the disk formation was already completed,  and the shape of the H$\alpha$ line confirms a close to pole-on view of the disk. We detect clear variations in the absolute values of the intensities, suggesting a highly dynamical evolution of the disk, which is also supported by the variations seen in the velocities of the emission peaks. However, the lack of detailed spectral monitoring prevents us from drawing a more solid conclusion. 

Multi-band photometric observations covering major outbursts in  Be stars are exceptionally rare but provide meaningful insight into the structure and evolution of the associated viscous decretion disks. As was shown in Figs.\,\ref{fig:colorcolor} and \ref{fig:colorcolorJapan}, the colour-magnitude and colour-colour diagrams of \HDtwentyfive\ trace disk formation and dissipation episodes, provide information on the disk density and viscosity, as well as on the inclination angle \citep{Haubois2012, Figueiredo2025}. Therefore, our  data represent a valuable resource for detailed future modelling of the disk and for investigating its physical properties.

\subsection{Quiescence of PY\,Gem}
\label{sect:quiescence}

In the long-term photometric light curve of PY\,Gem, we detect a contradictory pattern (Figure~\ref{mmt}). The star's  well defined photometric variability is only recorded until approximately 2008. Thereafter, the star seems to be in a continuous phase of quiescence with only low-amplitude variability. This is confirmed by the high-cadence \tess\ observations, which fall all into this quiet phase of the star. 

On the other hand, our spectroscopic data covering this quiet phase of PY\,Gem (from 2009 on) show clear variability in the H$\alpha$ intensity (i.e. equivalent width, see Figure~\ref{fig:HaHD}) and the velocities of the blue and red peaks (Table\,\ref{tab:PYGem}), implying a highly dynamical line-forming region, while the ratio of the emission peaks does not change significantly. An increase in the equivalent width indicates that more matter is located in the vicinity of the star. The subsequent decrease in equivalent width (sometimes within a couple of days, as seen in March 2021) implies that the region close to the stellar surface is being evacuated and the matter is either dissipated or re-accreted to the stellar surface. This (maybe periodic or at least cyclic) behaviour of matter supply and dissipation occurs without any significant change in the stellar brightness. According to the theoretical predictions by \citet{Haubois2012}, this could indicate a high inclination of the disk of $\sim 70\degr$.

The last spectrum from 2025 shows a significantly weaker double-peaked emission profile of the H$\alpha$ line, whereas the emission in \HeI$\lambda 6678$ disappeared completely (see Figure~\ref{fig:HaHD}) revealing a pure, rotationally broadened photospheric profile. This could suggest that the matter supply from the star has stopped and the decretion disk might dissipate completely (although it might also be just an observational bias due to the sparse spectroscopic sampling).

\subsection{Frequency analyses of the \tess\ light curves}
\label{sect:freq_anal}

The physical mechanism responsible for the Be phenomenon remains not completely understood.  Since stellar pulsations are thought to play an important role in triggering mass ejection and disk formation, the analysis of TESS photometry offers a valuable opportunity to probe the internal structure of Be stars and better understand the nature of this phenomenon.

The frequency spectra of high-cadence light curves of Be stars can display four different features. These are (i) long-term trends with timescales $> 2$\,days which are usually associated with flickers or episodes of mass ejections, (ii) low-frequency stochastic variations,
(iii) isolated frequencies, and (iv) frequency groups consisting of several closely spaced frequencies \citep[for more details, see e.g.,][]{2022AJ....163..226L}. The occurrence of frequency groups can be linked to the rapid rotation of the star \citep[see e.g.,][]{2018MNRAS.477.2183S}, and  frequency groups are seen in about 75\% of all Be stars \citep[including $\gamma$~Cas;][]{2021MNRAS.502..242L}. Those groups with low frequencies ($< 4$\,d$^{-1}$) are typically found to be g-mode pulsations, whereas higher frequency signals are consistent with p-modes \citep{RiviniusCarciofi2013, 2009A&A...506...95H, 2016A&A...588A..56B}. 


To investigate the frequency spectra of our objects, we combined the \tess\ light curves taken in consecutive sectors (i.e. sectors 43, 44, and 45 for both stars and sectors 71 and 72 for \HDtwentyfive).
The frequency analysis was performed using an iterative pre-whitening procedure implemented in {\sc Pyriod} \citep{Bell2020, Bell2022}. A total of 43 and 48 frequencies were initially extracted for PY\,Gem and HD\,253659, respectively. To avoid including unresolved peaks, frequencies separated by less than $1/T$, where $T$ is the time span of the observations, were considered indistinguishable, and only the component with the highest amplitude was retained. The multi-sinusoidal fit was then repeated using this reduced set of frequencies. Statistical significance was assessed using a threshold of four times the local noise level in the residual periodogram. The local noise was estimated as the mean residual amplitude within a sliding window of width $2$\,d$^{-1}$, evaluated in steps of $2$\,d$^{-1}$. Only frequencies with $S/N>4$ were retained for further analysis. The same $1/T$ criterion was also adopted when identifying harmonic and combination frequencies.  The identified frequencies are listed in Tables~\ref{tab:tessPYGem} and \ref{tab:tessHD25} for PY~Gem and HD\,253659, respectively.

The temporal evolution of the frequency spectrum was investigated using the weighted wavelet Z-transform (WWZ; \citealt{Foster1996}) and the Python package \texttt{libwwz}\footnote{ https://github.com/ISLA-UH/libwwz}. This method combines time and frequency information, enabling the detection of transient oscillations and temporal changes in both frequency and amplitude. The resulting WWZ scalograms provide a two-dimensional representation of the variability, making them particularly useful for studying the non-stationary behaviour commonly observed in Be stars.

\subsubsection*{\it PY\,Gem}

Starting with PY\,Gem, we distinguish four main groups of frequencies in the combined sectors $43-45$. We label these groups as $g1$ to $g4$ in Table~\ref{tab:tessPYGem}, following the notation of \citet{2021MNRAS.502..242L}. As observed in many Be stars, $g2$ is located at approximately twice the frequency of $g1$ \citep[e.g.,][]{2022AJ....163..226L}. Moreover, $g3$ and $g4$ have about twice and three times the frequency of $g2$, which is also similar to what was found for $\gamma$~Cas \citep{2021MNRAS.502..242L}. The first three groups span the typical range of g-mode pulsations whereas $g4$, which is centered at $\sim 7$\,d$^{-1}$, fits better to the range of p-modes. A frequency similar to the dominant one of $g4$ was also reported by \citet{2017AJ....153..252L} who identified it in the light curve of PY\,Gem from the KELT photometric survey.

The remaining frequencies appear as individual frequencies and are marked with $f_{0} - f_{9}$ in {Table~\ref{tab:tessPYGem}}. Taking a closer look at these, we identify five combination frequencies ($f_3=g4_3-g1_8$, $f_5=2f_1-2g1_9$, $f_6=3g1_3+g4_2$, $f_8=2f_4-g3_2$ and $f_9=3f_7-3g3_4$), leaving us with four individual frequencies ($f_1$, $f_2$, $f_4$ and $f_7$) that might be identified as additional p-mode pulsations. Such high-frequency signals are less common in Be stars and are typically found in $\beta$~Cephei pulsators. PY\,Gem might hence be classified as belonging to the class of Be stars that are $\beta$~Cephei hybrid pulsators \citep[similar to the Be star HD\,49330, see][]{2009A&A...506...95H}. Only about $10-20\%$ of the sample (independent of whether these are early, mid, or late-type Be stars) studied by \citet{2022AJ....163..226L} shows such isolated high frequencies, suggesting that hybrid pulsators are less common among Be stars and $\gamma$\,Cas analogs.

Sector 71 displays the same frequency groups. It also displays the frequency $f_0$ that we have not discussed yet. However, we find for $f_0$ a frequency of 0.4074\,d$^{-1}$, which is slightly higher than the value of 0.3924\,d$^{-1}$ found in the combined sectors.  Additionally, the frequency $g4_3$ has reduced its amplitude below the detection threshold in sector 71. This shift in frequency and amplitude changes have prompted us to examine the individual sectors more closely to check whether these observations might be real or an artifact, because of the single-sector observation with different cadence. 

\begin{figure}
  \centering
  \includegraphics[width=1\columnwidth]{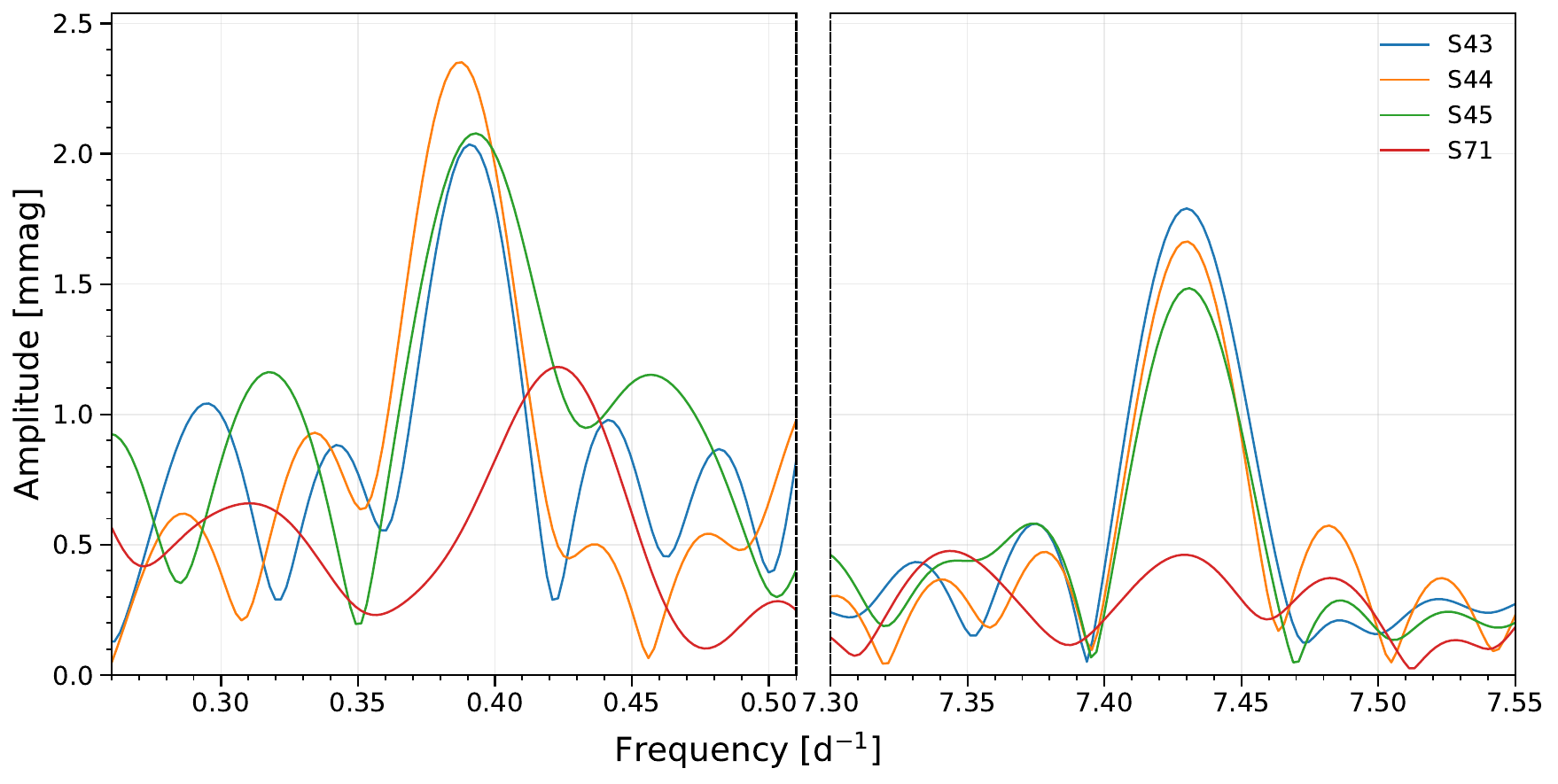}
  \caption{ Comparison of the FFTs of the individual sectors of PY\,Gem centred at $f_0=0.3923$\,d$^{-1}$ (left) and $g4_3=7.4297$\,d$^{-1}$ (right).} 
  \label{fig:FFT_freq_PYGem} 
\end{figure}
\begin{figure*}
  \centering
  \includegraphics[width=2\columnwidth]{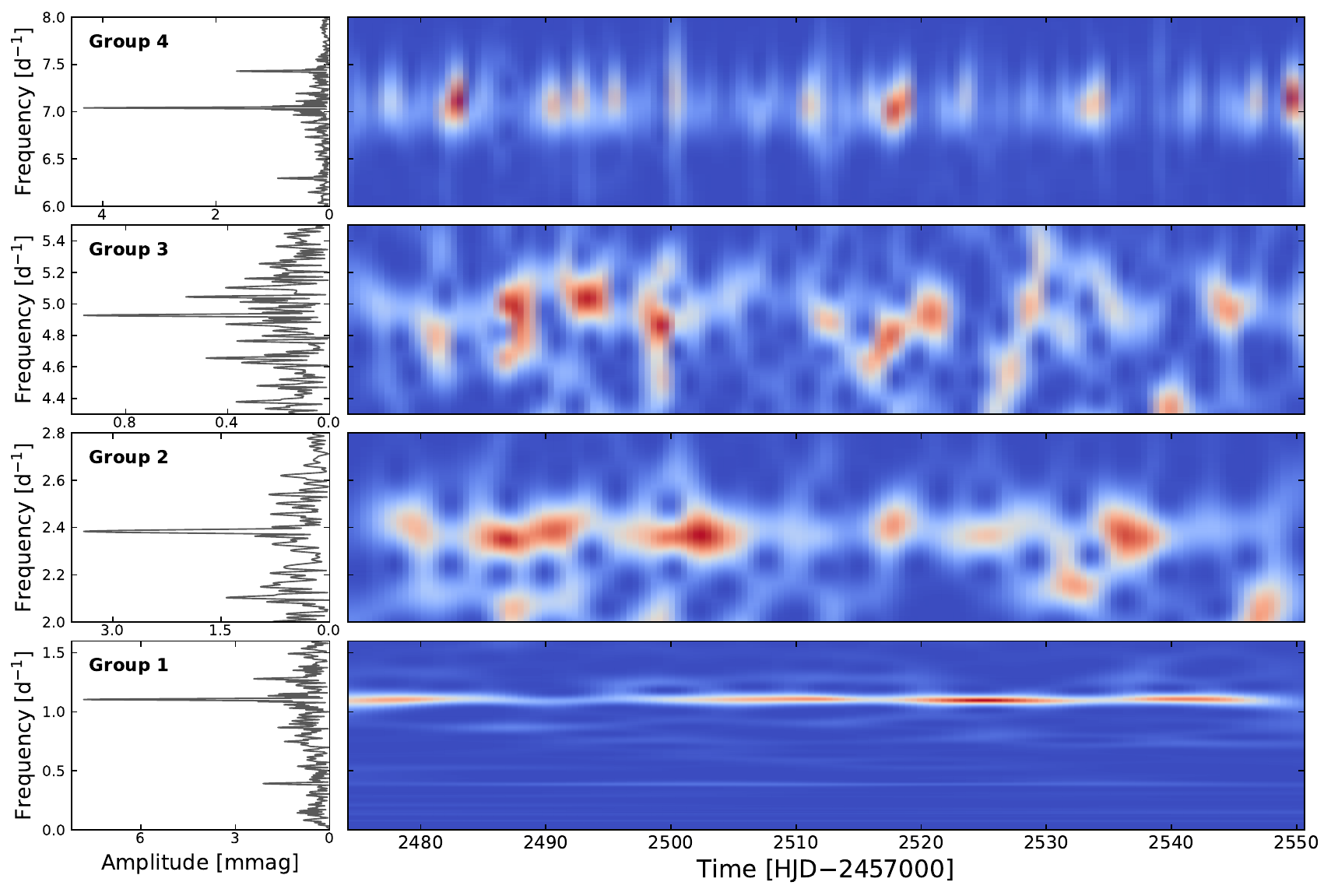}
  \caption{
  Amplitude spectra (left panels) and corresponding   scalograms (right panels) for the four identified frequency groups (groups~1--4, from bottom to top) for the combined sectors $43-45$ of PY\,Gem, normalized to the maximum power within each group. 
  } 
  \label{fig:wwz_PYGem} 
\end{figure*}
 
Figure~\ref{fig:FFT_freq_PYGem} shows the Fast Fourier Transform (FFT) centred at  $f0$ and $g4_3$ for each individual sector. We find that $f_0$ has slightly different values in each of them. In addition, its amplitude changes from sector to sector. The frequency $f_0$ might be a representative of $g0$, which is an additional frequency group in the domain $< 0.5$\,d$^{-1}$ often seen in Be stars. However, in that case, its frequency should remain constant. Instead, the incoherent frequency $f_0$ could be interpreted as the so-called \v{S}tefl frequency. In their studies of spectroscopic time series, \citet{1998ASPC..135..348S} found that during phases of temporarily enhanced H$\alpha$ emission (outbursts), spectral lines that are formed in layers above the photosphere display variability with a frequency not related to stellar pulsations and with projected rotation velocities in agreement with a location above the photosphere. Later, \citet{2016A&A...588A..56B} have proven that the same frequency can also be detected in the light curve of the star. Furthermore, these authors have shown that the \v{S}tefl frequency changes both its frequency and amplitude over time, in agreement with our findings for $f_0$. They also proposed that for the detection of the \v{S}tefl frequency in photometry, the system must be seen under a high inclination angle, which was  confirmed by numerical studies \citep{2026arXiv260705270R}. The presence of the \v{S}tefl frequency in PY~Gem agrees with the high inclination suggested earlier.  

As mentioned in Sect.~\ref{sect:quiescence}, PY\,Gem is in a sort of quiescence state since 2008 with no indication of outbursts. Still, it is very actively pulsating with at least four frequency groups detected. Moreover, its cyclic (or periodic) changes in its emission spectrum speak in favor of a highly dynamical line-forming region of H$\alpha$ in the sense that matter is supplied cyclically and then dissolves, leading to an increase and a subsequent decrease in the equivalent width of the line. This region, which marks the interface between the star and its disk, is also the region in which the \v{S}tefl frequency is excited, for which \citet{2016A&A...588A..56B} speculate that the mass-loss process could be responsible. 

For frequency $g4_3$, interpreted as a p-mode, we also noticed unstable behaviour. This frequency displays similar values in sectors 43, 44, and 45, with a gradual decrease in amplitude (see Figure~\ref{fig:FFT_freq_PYGem}). However, in Sector 71 it is no longer detected.  The spectra obtained during sector 44 (October 2021), display a prominent double-peak H$\alpha$ emission, with normalised peak intensities of approximately 1.43 to 1.53. These values are among the highest measured during the monitoring and are clearly larger than the 1.21 peak intensity observed shortly before sector 71, in September 2023. This suggests that the pulsation amplitude of this mode may be modulated by changes in the outer layers of the star associated with disk feeding, while the \v{S}tefl frequency may independently trace the variability at the star-disk interface.

\begin{figure*}
  \centering
  \includegraphics[width=2\columnwidth]{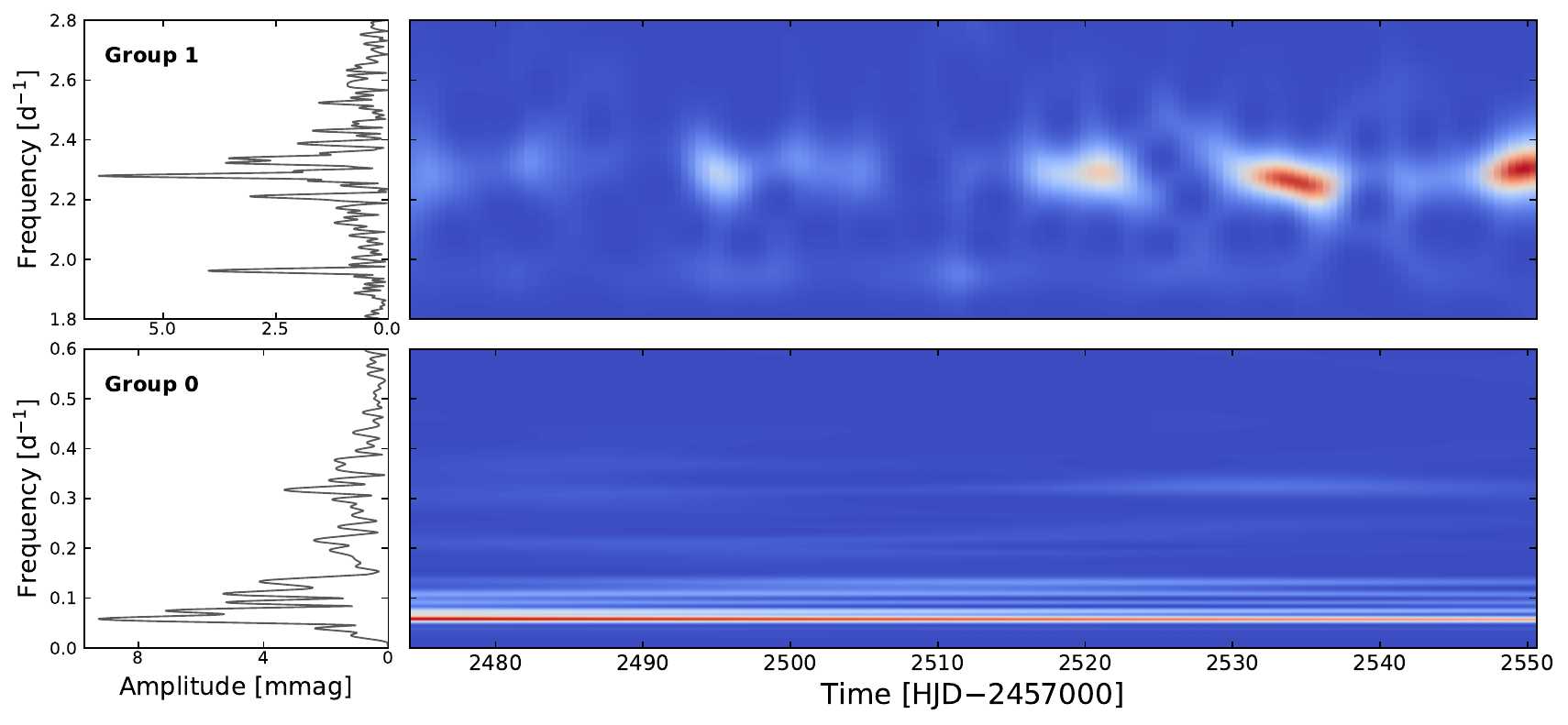}
  \caption{Same as Figure~\ref{fig:wwz_PYGem} but for \HDtwentyfive. }
  \label{fig:wwz_HD25} 
\end{figure*}

Our frequency analysis also revealed numerous linear combinations of independent frequencies. Among them, three stand out: $g1_7=g4_2-f1$, $g2_4=f_7-g4_3$ and $g1_{10}=g2_3-g1_5$. For these cases, the amplitude of the ‘child’ frequency exceeds the sum of the amplitudes of its ‘parents’ modes. This amplification may indicate nonlinear coupling between pulsation modes rather than a simple linear beating. In Be stars,  this interaction is particularly relevant to mass loss, through which coupled modes may concentrate pulsation energy and angular momentum at the stellar surface, periodically facilitating the ejection of material into the circumstellar disk. A similar case was reported for $\eta$ Cen, where a difference frequency associated with two non-radial pulsation modes reached approximately three times the sum of the parent-mode amplitudes and was linked to enhanced mass loss \citep{2016A&A...588A..56B}. These three amplified combinations found could then trace pulsation-driven mass loss processes, although accidental numerical matches cannot be excluded.

Figure \ref{fig:wwz_PYGem}  displays the amplitude spectra and scalograms for the four identified frequency groups in PY\,Gem, showing two distinct types of variability. The lowest-frequency group displays a more or less stable behaviour over the observing period, whereas the higher-frequency groups exhibit pronounced temporal changes with patchy patterns possibly due to beating phenomena of neighbouring frequencies within the densely populated groups \citep{2022AJ....163..226L}.

\subsubsection*{\it \HDtwentyfive}

In the combined sectors $43-45$, \HDtwentyfive\  displays two main groups of frequencies in the low-frequency range (see Figure~\ref{fig:wwz_HD25}). The group with 
frequencies lower than 0.65\,d$^{-1}$, we identify as $g0$, and the second group in the range $1.9-2.{8}$\,d$^{-1}$ as $g1$. We also found several combination frequencies within each group. Among the isolated frequencies, only $f_2$ is independent, suggesting that this frequency is a $p$-mode. Additionally, an amplified difference frequency ($g0_6=g1_{12}-g1_7$) was also found in HD\,253659, indicating a possible nonlinear mode coupling and suggesting that the same pulsation-driven mass-loss mechanism may be operating in this object.

\begin{figure}
  \centering
  \includegraphics[width=1\columnwidth]{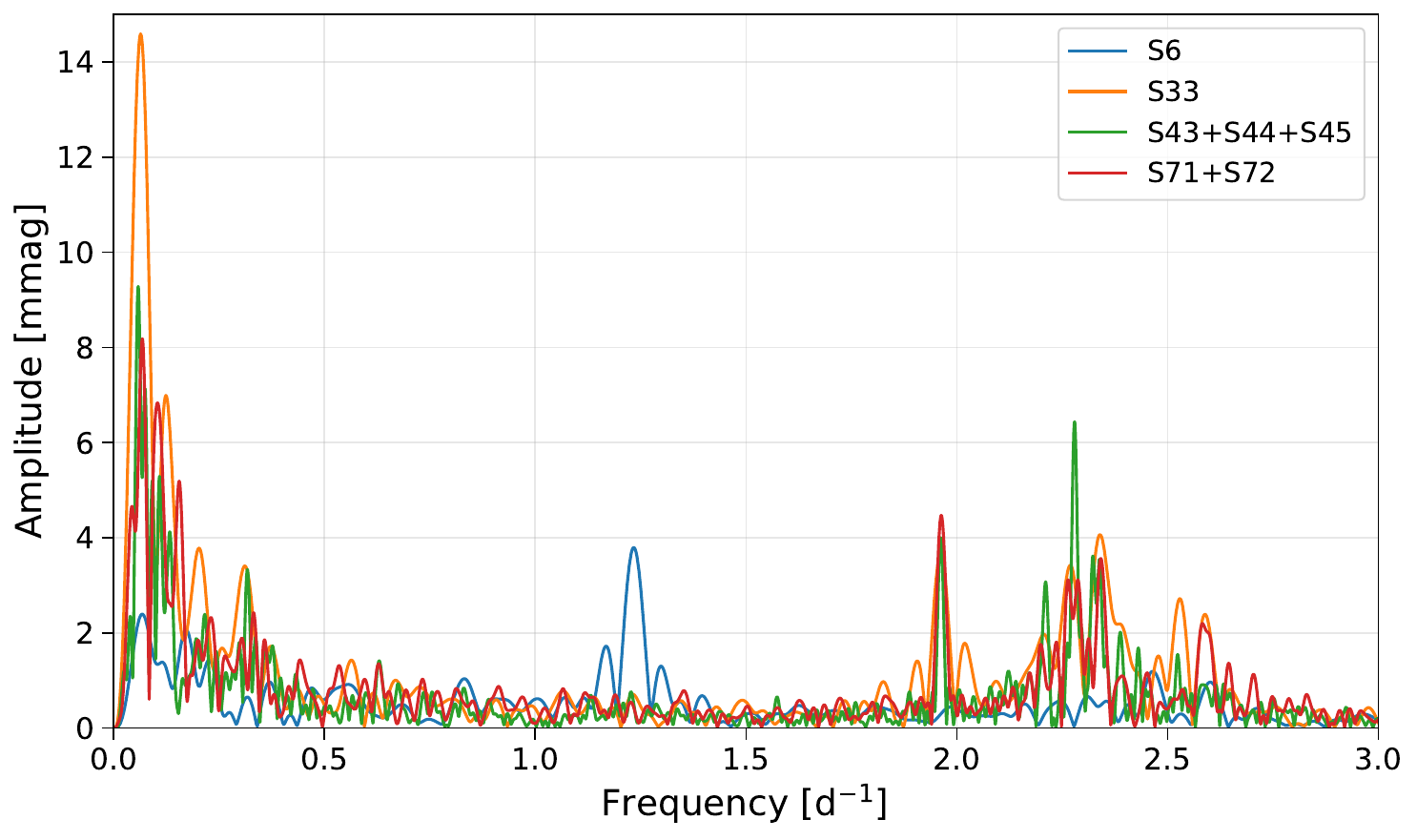}
  \caption{Comparison of the FFTs of individual (6, 33) and combined sectors ($43-45$, $71-72$) of HD\,253659.}
  \label{fig:FFT_freq_HD25} 
\end{figure}

Figure \ref{fig:FFT_freq_HD25} shows a comparison of the FFTs for individual and combined sectors, showing that the same frequency pattern is found in sector 33 and in the combined sectors $71+72$ with slight modulation of the relative amplitudes of the individual frequencies. However, inspecting sector 6, the situation is different. Sector 6 lacks the signal at 1.96\js{2}\,d$^{-1}$ (identified as $g1_1$ in the later sectors), while the remaining pattern of group $g1$ appears to be slightly shifted to higher frequencies. In addition, it shows the strongest signal at 1.234\,d$^{-1}$. This frequency is absent in the observations of the later sectors. 

To better understand the changes in the amplitude spectra of the different \tess\ light curves over time, we inspect the portions of the major outburst to which their observing periods correspond (see Figs.\,\ref{kws} and \ref{kws_tess}). Sector 6 was observed in the middle of the major outburst and during the third brightening, i.e. the disk build-up phase. Sector 33 was obtained during the dissipation (or recovery) phase of the major outburst, while sectors $43-45$ fall into the subsequent (relatively) quiescent phase. The sectors $71-72$ also describe a recovery phase after the next (moderate) outburst. 

In a study of continuous photometric monitoring of an entire moderate outburst in the early-type Be star HD\,49330 with the CoRoT satellite, \citet{2009A&A...506...95H} found compelling evidence for changes in the Fourier spectrum over the event. In particular, a gradual emergence of frequency groups and individual frequencies during the precursor and outburst phases, which reach their maximum amplitude during the outburst. At the same time, frequencies that are prominent during quiescence are reduced. During the relaxation phase, the Fourier spectrum seems to recover to the pre-outburst stage. In their study, \citet{2009A&A...506...95H} attribute the appearance of transient frequencies to rapid changes in the structure of the outer layers of the star during the outburst.

Our \tess\ data cover only small and individual portions of the events seen in the ground-based light curve of \HDtwentyfive\, but the trend observed in our Fourier spectra is very similar. 
However, we also note a significant difference compared to HD\,49330. The de-trended \tess\ light curves of all sectors but sector 6 show clear indications for recurrent flickers. These events with cycles of $10-20$\,d are most likely responsible for the high-amplitude low-frequency signals ($<0.4$\,d$^{-1}$) in the corresponding Fourier spectra (which we tentatively labeled as group $g0$).  

 Similarly to PY\,Gem, the scalograms for \hdtwentyfive\ (Figure~\ref{fig:wwz_HD25}) reveal two distinct types of variability. The low-frequency signal (Group 0) remains coherent throughout the observations, whereas the higher-frequency signal (Group 1) exhibits pronounced temporal variations in amplitude and a lower degree of coherence, possibly connected to a beating phenomena.

\section{Discussion}\label{sec:discussion}

Our investigation of \HDtwentyfive\ and PY\,Gem was originally motivated by the discovery of nebulae surrounding them. But our current study presented above, including the determination of the positions of the objects in the HR diagram, reveals no unusual stellar properties that could naturally explain the origin of the nebulae. Thus, this raises the question of a possible connection between the formation of the observed circumstellar structures, the evolution of the decretion disk, and the variability properties of the central star. Since pulsations have been proposed as one of the possible mechanisms facilitating the ejection of material  \citep[see e.g.][]{ 2003A&A...411..229R,Cranmer2009, RiviniusCarciofi2013, Baade2017, Semaan2018}
and its transfer into the circumstellar environment, it is particularly interesting to investigate whether changes in the pulsational activity of the star are related to episodes of disk formation and dissipation. Therefore, we investigated the photometric variability of the stars on different timescales, combining long-term light curves, which trace large-scale changes in the circumstellar disks, with high-cadence \tess\ photometry to study short-term variability and its frequency structure.

Despite being in a similar evolutionary state, the objects differ significantly in their photometric appearance. While on long time scales \HDtwentyfive\ exhibits  quasi-periodic minor outbursts with low duty cycles followed by a high-amplitude major outburst event that lasted about six years  with several disk formation and dissipation phases, PY\,Gem has ceased its typical $\gamma$~Cas-type variability and instead displayed only small-amplitude variations over the past 20~years. 

The high-cadence \tess\ data clearly reveal the intrinsic properties of the targets and highlight  differences between them. The frequency content observed in PY\,Gem remains relatively stable from sector to sector.  With both $g$ and $p$-mode pulsations, PY\,Gem might belong to the group of Be stars that are $\beta$~Cephei hybrid pulsators. The object also shows an incoherent low-frequency signal, which we tentatively classify as a \v{S}tefl frequency originating from the interface region between the star and its disk \citep{2016A&A...588A..56B}. Support for such an interpretation is provided by the cyclic (or periodic) variability of emission lines like H$\alpha$ and \HeI that are formed in the same region. In contrast,  \HDtwentyfive\ exhibits substantial changes in its variability pattern, with the frequency structure undergoing significant reorganization during different phases of its activity.
Its \tess\ observations reveal  numerous $g$-modes and one $p$-mode as well as low-frequency variability that might be related to flickers and hence to possible density variations in the transition region between star and disk respectively the innermost disk region.  Unfortunately, our spectroscopic coverage is insufficient for more solid conclusions. The individual sectors map different phases of the major outburst, and the general trend of the variabilities seen in the amplitude spectra agree well with those reported during outburst cycles  \citep[see e.g.][]{2009A&A...506...95H}.

Nevertheless, the mass-loss and disk-forming activity detected in \HDtwentyfive\ could indicate that (some or certain) Be stars have the potential to create dusty nebulae. Support for this is provided by the SED of \HDtwentyfive, which contains a warm ($\sim550$\,K) dusty component. Inspection of near-infrared images unveils no clear evidence of extended emission, so that this warm dust must be compact. Further support for high activity of Be stars that can lead to an accumulation of circumstellar material with conditions favouring the formation of molecules is provided by 12\,Vul. This Be star was proposed by \citet{2021A&A...647A.164C} to be in a transition phase between classical Be stars and B[e] supergiants, which are objects surrounded by dense molecular and dusty disks \citep[see, e.g.,][]{2019Galax...7...83K}.
As such, outbursts and pulsation-triggered mass ejections from Be stars might also be related to instabilities similar to those recently proposed for B[e] supergiants \citep{2024ApJ...963..131N}.

The presumed lower mass-loss activity of PY\,Gem, which could be inferred from the absence of strong photometric variations, might simply be a misinterpretation of the actual conditions. As numerical computations have shown \citep{Haubois2012}, the detection of variability strongly depends on the inclination angle of the system. PY\,Gem might have an orientation ($\sim 70\degr$) that suppresses the detection of (strong) photometric variability, while its activity is still traced in spectroscopy.

\section{Conclusions}\label{sec:conclusions}

We studied the objects PY\,Gem and \HDtwentyfive\ associated with nebulae that were discovered during the search for evolved massive stars with circumstellar structures in infrared images.  Detailed study of the stars showed that they do not belong to evolved massive stars and are not related to the LBV phase, during which the formation of circumstellar nebulae is more natural. Instead, our analysis revealed that both stars are classical Be-stars, which have similar masses and luminosities. 

The long-term light curves constructed from photometric data collected from various archives revealed the outburst activity of HD\,253659 with major outburst between 2016 and 2022 and, in contrast, the prolonged quiescent state of PY\,Gem. For \HDtwentyfive\  we traced  disk formation and dissipation episodes in colour-magnitude and colour-colour diagrams. Collected data represent a valuable resource for detailed future modelling of the disk and for investigating its physical properties. 

The high-cadence \tess\ observations reveal markedly different variability patterns in the two stars. PY\,Gem shows relatively stable frequency content across the observed sectors, with both g- and p-mode pulsations suggesting that it may be a $\beta$ Cephei hybrid pulsator; its additional low-frequency signal may represent a \v{S}tefl  frequency associated with the star--disk interface. In contrast, HD\,253659 exhibits substantial reorganization of its frequency spectrum between different phases of the major outburst, including numerous g-modes, a p-mode, and low-frequency variability that may be related to density variations in the star--disk transition region or the innermost disk. These results suggest a close connection between the changing pulsational/photometric variability and the evolution of the circumstellar disk during Be-star activity cycles.

All spectral lines observed in \HDtwentyfive\ are in emission, whereas the spectrum of PY\,Gem contains absorption lines, which allowed us to determine its projected rotational velocity,  $v~\sin{i}=303\pm15~$\kms. Spectral monitoring of both stars did not detect any significant changes in radial velocity and thus provided no evidence for binarity.

Our investigation of these objects was originally motivated by the discovery of nebulae surrounding them. To conclude,  we suggest that the nebulae detected around \HDtwentyfive\ and PY\,Gem may be linked to the Be phenomenon itself. Testing this hypothesis requires a dedicated statistical analysis of a larger sample of Be stars and will be the subject of a forthcoming paper.

\section*{Acknowledgements}\label{sec:acknow}

We dedicate this publication to the memory of our dear colleague Vasilii Gvaramadze, who passed away on September 2nd, 2021. This is another article that would not have been possible without his idea, both studied stars are objects selected by him. The search for evolved stellar objects by means of the detection of their circumstellar nebulae in modern sky surveys was a topic that Vasilii successfully developed for the last years of his life. 
\\
We thank the anonymous referee for constructive feedback that improved the manuscript. 
We acknowledge support from the Czech Science Foundation (project GA\v{C}R 25-17532S). 
This work was co-funded by the European Union (Project 101183150 – OCEANS) and supported by the Czech Ministry of Education, Youth and Sports (MEYS) (Project No. CZ.02.01.01/00/22\_008/0004632 -- FORTE). This research is partially based on the data acquired by the Perek 2m telescope at the Astronomical Institute of the Czech Academy of Sciences (CAS) in Ond\v{r}ejov and the robotic telescope FRAM-ORM, which are supported by the project RVO:67985815  of the CAS and by the the grant of the MEYS LM2018102, respectively. This work has made use of: the BeSS database, operated at LESIA, Observatoire de Meudon, France (http://basebe.obspm.fr); data products from the Wide-field Infrared Survey Explorer (WISE), which is a joint project of UCLA and JPL/Caltech, funded by NASA; and data from the TESS mission, publicly available from the Mikulski Archive for Space Telescopes (MAST). This research has also made use of the SIMBAD database and the VizieR catalogue access tool, operated at CDS, Strasbourg, France.







\section*{DATA AVAILABILITY}
The data underlying this article will be shared on reasonable request to the corresponding author.

\bibliographystyle{mnras}
\bibliography{Nebulae}

\appendix

\section{Photometric monitoring with ground-based robotic telescopes and space missions}

   \begin{figure*}
   \centering
   \includegraphics[width=2.0\columnwidth,clip]{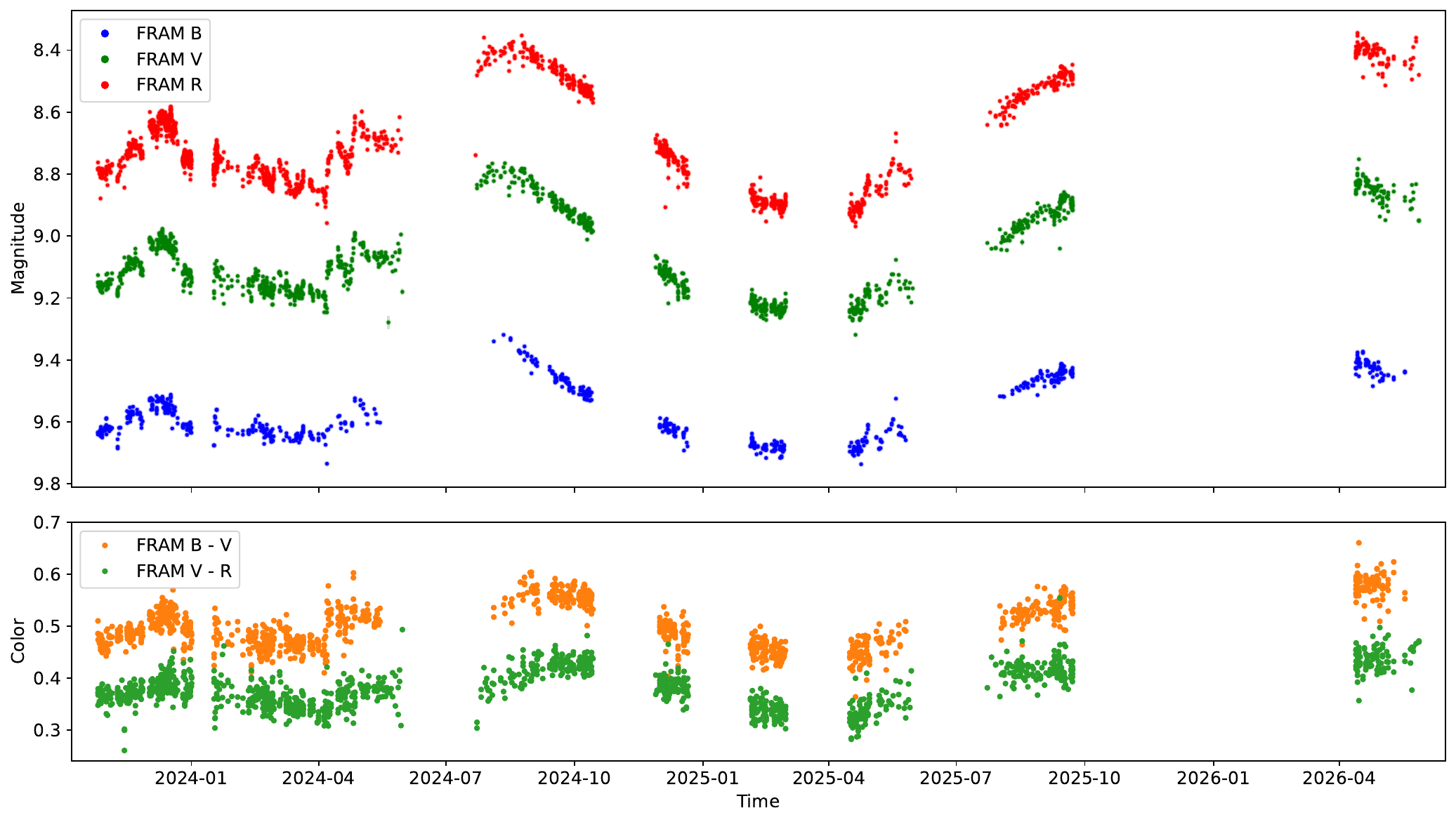}
   \caption{{\it Upper panel:} light curves of \HDtwentyfive \ in the $B$ (blue dots), $V$ (green dots), and $R$ (red dots) bands based on the FRAM multi-band photometry.  {\it Lower panel:} behaviour of the corresponding colours.  \label{fram}}
   \end{figure*}

\section{Spectroscopic data}
\begin{table}\centering
\caption{Log of observations at the Perek 2-m telescope using the Coude and OES spectrographs, along with the dates of observations and instruments of the spectra from the BeSS database that we used for spectral monitoring.
R is spectral resolution, SMM is Santa Maria de Montmagastrell, UH is University Hills (Los Angeles), DW is Desert Wing, SJ is San Jose,  ASJ is observatoire Antibes Saint-Jean, MR is Mill Ridge \label{tab:log11} }
\begin{tabular}{lcccc}
\hline
     & Date        &    Exp [s]     &\multicolumn{1}{c}{Instrument}  & R \\
\hline
{\multirow{15}{*}{\rotatebox[origin=c]{90}{PY\,Gem}}} 
    &  2009-03-17      &  2428   & BeSS/PIERA   &  6000 \\ 
    &  2012-03-10      &  1818   & BeSS/SMM  &  5000\\ 
    &  2013-04-11      &  3012   & BeSS/PIERA  & 5866  \\ 
    &  2015-03-01      &  3684   & BeSS/MR  & 8756 \\ 
    &  2017-03-19      &  3624   & BeSS/MR  & 8810 \\
    &  2019-12-16      &  7257   & BeSS/UH  & 10975 \\ 
    &  2020-04-10      &  2794 & Coude H$\alpha$ & 13000  \\
    &  2020-04-12      &  4801 & Coude H$\alpha$ & 13000 \\
    &  2020-04-26      &  3601 & Coude H$\alpha$ & 13000 \\
    &  2021-03-21      &  12045   & BeSS/DW  & 15917 \\ 
    &  2021-02-04      &  7200 &   OES &  50000\\
    &  2021-03-25      &  8202 & Coude H$\alpha$ & 13000 \\
    &  2021-10-25      &  4601 & Coude H$\alpha$ & 13000 \\
    &  2021-10-28      &  1582 & Coude H$\alpha$ & 13000 \\
    &  2021-10-30      &  1063 & Coude H$\alpha$ & 13000 \\
    &  2022-01-06      &  4801 & Coude H$\alpha$ & 13000 \\
    &  2023-01-24      &  3600 & BeSS/SJ  & $-$ \\ 
    &  2023-09-24      &  2794 & Coude H$\alpha$ & 13000 \\
    &  2023-10-07      &  4801 & Coude NIR & 13000 \\
    &  2024-04-08      &  4801 & Coude H$\alpha$ & 13000 \\
    &  2025-01-19      &  1650 & Coude H$\alpha$ & 13000 \\
  \hline    
{\multirow{9}{*}{\rotatebox[origin=c]{90}{HD\,253659~}}}
    &  2020-09-21      & 3001  & Coude  H$\alpha$& 13000\\ 
    &  2020-11-06      & 2407  & BeSS/ASJ   & 12000\\   
    &  2021-03-26      & 3601  & Coude H$\alpha$ & 13000\\   
    &  2021-03-26      & 3601  & Coude NIR  & 13000\\ 
    &  2021-04-04      & 4001  & Coude NIR & 13000\\ 
    &  2021-10-25      & 4801  & Coude H$\alpha$ & 13000\\  
    &  2023-10-07      & 5201  & Coude H$\alpha$ & 13000\\ 
    &  2024-04-08      & 5201  & Coude H$\alpha$ & 13000\\
    &  2024-11-09      & 5201  & Coude H$\alpha$ & 13000\\
    & 2025-03-02       & 5201  & Coude H$\alpha$ & 13000\\
    & 2025-03-02       & 4001  & Coude NIR & 13000\\  
    & 2025-03-12       & 4206  & Coude H$\alpha$ & 13000\\
\hline
\end{tabular}
\end{table}

\begin{table*}\centering
\caption{Parameters of the H$\alpha$ line in the spectra of PY\,Gem. $Vb$ and $Vr$ are the velocities of the blue and red peaks, correspondingly, $Va$ is the one of the central absorption. $Ib$, $Ia$ and $Ir$ are their relative intensities.  \label{tab:PYGem} }
\begin{tabular}{lcc ccc cc}
\hline
\multicolumn{1}{c}{\multirow{2}{*}{Date}} & \multirow{2}{*}{JD-2400000}            &   $Vb$ &  $Va$   &   $Vr$    &   \multirow{2}{*}{$Ib$}  &  \multirow{2}{*}{$Ia$}     &   \multirow{2}{*}{$Ir$}  \\
                         &  & [\kms] &  [\kms] & [\kms]    &       &         &       \\ 
\hline
2009-03-17 &  54908.4 &  $-158.5\pm22.8$ & $  1.3\pm22.8$ & $161.2\pm22.8$   &  1.32  &   1.00 & 1.34  \\
2012-03-10 &  55997.5 &  $-226.4\pm22.8$ & $-20.8\pm22.8$ & $184.7\pm22.8$   &  1.35  &   0.95 & 1.38  \\
2013-04-11 &  56394.4 &  $-180.5\pm22.8$ & $  2.2\pm22.8$ & $184.9\pm22.8$   &  1.40  &   1.01 & 1.43  \\
2015-03-01 &  57082.7 &  $-178.7\pm22.8$ & $  4.0\pm22.8$ & $186.7\pm22.8$   &  1.20  &   0.88 & 1.20  \\
2017-03-19 &  57831.6 &  $-204.3\pm22.8$ & $  1.2\pm22.8$ & $206.8\pm22.8$   &  1.36  &   0.92 & 1.35  \\
2019-12-16 &  58833.9 &  $-191.5\pm22.8$ & $ 14.0\pm22.8$ & $196.8\pm22.8$   &  1.38  &   0.98 & 1.44  \\
2020-04-10 &  58950.3 &  $-171.8\pm10.5$ & $ 18.0\pm10.5$ & $176.3\pm10.5$   &  1.33  &   0.99 & 1.36  \\
2020-04-12 &  58952.3 &  $-180.3\pm10.5$ & $ 20.1\pm10.5$ & $178.3\pm10.5$   &  1.35  &   0.97 & 1.37  \\
2020-04-26 &  58966.3 &  $-177.0\pm10.5$ & $ 23.5\pm10.5$ & $181.7\pm10.5$   &  1.38  &   0.99 & 1.41  \\
2021-03-21 &  59294.7 &  $-181.4\pm22.8$ & $  1.3\pm22.8$ & $184.0\pm22.8$   &  1.54  &   1.03 & 1.56  \\
2021-03-25 &  59299.3 &  $-169.7\pm10.5$ & $-22.0\pm10.5$ & $167.9\pm10.5$   &  1.32  &   0.94 & 1.35  \\
2021-10-25 &  59512.6 &  $-168.1\pm10.6$ & $  0.7\pm10.6$ & $169.5\pm10.6$   &  1.49  &   1.00 & 1.49  \\
2021-10-28 &  59516.4 &  $-167.9\pm10.6$ & $-30.8\pm10.6$ & $180.3\pm10.6$   &  1.43  &   1.01 & 1.48  \\
2021-10-30 &  59517.6 &  $-169.7\pm10.6$ & $ -0.9\pm10.6$ & $178.5\pm10.6$   &  1.52  &   1.01 & 1.53  \\
2022-01-06 &  59586.3 &  $-173.0\pm10.5$ & $ -4.2\pm10.5$ & $164.6\pm10.5$   &  1.54  &   1.03 & 1.52  \\
2023-01-24 &  59968.7 &  $-257.4\pm22.8$ & $ -6.2\pm22.8$ & $245.0\pm22.8$   &  1.19  &   0.81 & 1.17  \\
2023-09-24 &  60211.5 &  $-223.0\pm10.6$ & $-22.5\pm10.6$ & $241.3\pm10.6$   &  1.21  &   0.84 & 1.21  \\
2024-04-08 &  60410.0 &  $-245.5\pm10.5$ & $ -2.9\pm10.5$ & $260.7\pm10.5$   &  1.16  &   0.78 & 1.16  \\
2025-01-19 &  60695.4 &  $-254.4\pm10.5$ & $ 19.9\pm10.5$ & $315.3\pm10.5$   &  1.09  &   0.77 & 1.11  \\
\hline
\end{tabular}
\end{table*}
\begin{table*}\centering
\caption{Parameters of the H$\alpha$ line in the spectra of \HDtwentyfive. $Vb$ and $Vr$ are velocities of the blue and red peaks, respectively, $Va$ is the one of the central absorption. $Ib$, $Ia$ and $Ir$ are their relative intensities, and $EW$ is the equivalent width. \label{tab:HD25} }
\begin{tabular}{lcc ccc cc ccc}
\hline
\multirow{2}{*}{Date}&\multirow{2}{*}{JD-2400000} & $V$ & $(V-I_c)$ &  $Vb$  &   $Va$  &   $Vr$   &  $Ib$   &  $Ia$  &  $Ir$ & $EW$ \\
            & &   [mag]         &   [mag]       & [\kms]     & [\kms]       &  [\kms]       &     & & & [\AA] \\
\hline
2020-09-21 & 59113.8 & 9.12 &     & $-45.1\pm10.5$ & $44.5\pm10.5$  &  $115.3\pm10.5$  &  1.29 & 1.26 & 1.30    & \\
2020-11-06 & 59159.6 & 9.15 &0.78 & $-64.1\pm10.5$ &  $35.9:\pm10.5$ &   $90.1\pm10.5$  & 2.68  & 2.54:&  2.69   & 15.79 \\ 
2021-03-25 & 59300.3 & 9.24 &     & $-50.0\pm10.5$ & $18.3\pm10.5$   &  $100.3\pm10.5$  &  2.55 & 2.29 & 2.60    &  14.59 \\ 
2021-10-25 & 59512.8 & 9.32 &0.52 & $-55.5\pm10.5$ & $40.1\pm10.5$   &  $122.7\pm10.5$  &  2.62 & 2.25 & 2.54    &  13.57 \\ 
2023-10-07 & 60224.8 & 9.10 &0.72 & $-60.0\pm10.5$ & $76.4\pm10.5$   &  $142.1\pm10.5$  &  1.85 & 1.59 & 1.83    &  9.51 \\ 
2024-04-08 & 60409.3 & 9.15 &     & $-60.5\pm10.5$ & $39.3\pm10.5$   &  $130.5\pm10.5$  & 2.01  & 1.69 &  1.94   & 8.32\\ %
2024-11-09 & 60624.4 & 9.05 & 0.78& $-30.8\pm10.5$ & $29.2\pm10.5$   & $129.3\pm10.5$   & 2.18  & 2.08 & 2.23    &13.21 \\ 
2025-03-02 & 60740.0 & 9.23 &     & $-92.0\pm10.5$ & $9.9\pm10.5$    &   $86.4\pm10.5$  &  2.51 & 2.24 & 2.54    & 13.40\\
2025-03-12 & 60750.0 & 9.23 &     & $-94.9\pm10.5$ & $-12.5\pm10.5$  &  $85.6\pm10.5$   & 2.3   & 2.02 & 2.38    & 11.46\\
\hline
\end{tabular}
\end{table*}

\section{\tess\ frequency tables}

\begin{table}
\caption{List of frequencies and amplitudes detected, along with their
errors, in the light curves of PY\,Gem in the combined \tess\ sectors
43, 44, and 45. Frequency groups are labelled following the notation of
\citet{2021MNRAS.502..242L} and \citet{2022AJ....163..226L}, while
isolated frequencies are labelled as $f$. 
\label{tab:tessPYGem}}
\centering
\begin{tabular}{cccccc}
\hline
\multirow{2}{*}{Id} &
Freq. &
$\sigma_{\rm Freq}$ &
Amp. &
$\sigma_{\rm Amp}$ &
Comb. \\
&
[d$^{-1}$] &
[d$^{-1}$] &
[mmag] &
[mmag] &
\\
\hline
$f0$        & 0.392301  & 0.000009 & 2.198678 & 0.002752 & \\
$g1_{1}$    & 0.747429  & 0.000015 & 1.493956 & 0.002831 & \\
$g1_{2}$    & 0.764372  & 0.000017 & 1.319377 & 0.002830 & \\
$g1_{3}$    & 0.865053  & 0.000015 & 1.292551 & 0.002777 & \\
$g1_{4}$    & 0.987502  & 0.000017 & 1.240231 & 0.002837 & \\
$g1_{5}$    & 1.047756  & 0.000017 & 1.220930 & 0.002870 & \\
$g1_{6}$    & 1.088357  & 0.000016 & 1.342233 & 0.002863 & \\
$g1_{7}$    & 1.104142  & 0.000003 & 7.791497 & 0.002871
             & $g4_{2}-f1$ \\
$g1_{8}$    & 1.136812  & 0.000015 & 1.450272 & 0.002900 & \\
$g1_{9}$    & 1.223356  & 0.000013 & 1.698407 & 0.002847 & \\
$g1_{10}$   & 1.278970  & 0.000008 & 2.490439 & 0.002816
             & $g2_{3}-g1_{5}$ \\
$g1_{11}$   & 1.493178  & 0.000016 & 1.231250 & 0.002764
             & $3g1_{9}-2g1_{6}$ \\
$g2_{1}$    & 1.935899  & 0.000019 & 1.036524 & 0.002775 & \\
$g2_{2}$    & 2.103877  & 0.000015 & 1.315188 & 0.002783
             & $f4-g4_{2}$ \\
$g2_{3}$    & 2.326436  & 0.000022 & 0.948608 & 0.002817 & \\
$g2_{4}$    & 2.383455  & 0.000006 & 3.405144 & 0.002816
             & $f7-g4_{3}$ \\
$g3_{1}$    & 4.175418  & 0.000063 & 0.334088 & 0.002929
             & $f2-2g1_{4}$ \\
$g3_{2}$    & 4.203657  & 0.000050 & 0.418726 & 0.002958 & \\
$g3_{3}$    & 4.655531  & 0.000043 & 0.459293 & 0.002942
             & $2f0+2g2_{1}$ \\
$g3_{4}$    & 4.870144  & 0.000044 & 0.458923 & 0.002997 & \\
$g3_{5}$    & 4.927180  & 0.000020 & 1.015686 & 0.003036
             & $f2-g1_{9}$ \\
$g3_{6}$    & 5.045448  & 0.000035 & 0.582539 & 0.003017
             & $f0+2g2_{3}$ \\
$g3_{7}$    & 5.102287  & 0.000040 & 0.495948 & 0.003009
             & $f2-g1_{5}$ \\
$f1$        & 5.935636  & 0.000058 & 0.338215 & 0.003068 & \\
$f2$        & 6.150013  & 0.000057 & 0.346288 & 0.003117 & \\
$f3$        & 6.295146  & 0.000022 & 0.890832 & 0.003120
             & $g4_{3}-g1_{8}$ \\
$g4_{1}$    & 6.895047  & 0.000068 & 0.288879 & 0.003214
             & $3g1_{1}+2g2_{3}$ \\
$g4_{2}$    & 7.039972  & 0.000005 & 4.295556 & 0.003238 & \\
$g4_{3}$    & 7.429757  & 0.000013 & 1.464165 & 0.003288 & \\
$f4$        & 9.142417  & 0.000102 & 0.189828 & 0.003641 & \\
$f5$        & 9.423250  & 0.000064 & 0.305623 & 0.003723
             & $2f1-2g1_{9}$ \\
$f6$        & 9.639517  & 0.000070 & 0.280176 & 0.003782
             & $3g1_{3}+g4_{2}$ \\
$f7$        & 9.813083  & 0.000053 & 0.369789 & 0.003832 & \\
$f8$        & 14.081260 & 0.000121 & 0.160916 & 0.006166
             & $2f4-g3_{2}$ \\
$f9$        & 14.829743 & 0.000131 & 0.148067 & 0.007017
             & $3f7-3g3_{4}$ \\
\hline
\end{tabular}
\end{table}

\begin{table}
\caption{Same as in Table \ref{tab:tessPYGem} for HD 253659.
\label{tab:tessHD25}}
\begin{center}
\begin{tabular}{cccccc}
\hline
\multirow{2}{*}{Id} &
Freq. &
$\sigma_{\rm Freq}$ &
Amp. &
$\sigma_{\rm Amp}$ &
Comb. \\
&
[d$^{-1}$] &
[d$^{-1}$] &
[mmag] &
[mmag] &
\\
\hline
$g0_{1}$  & 0.014067 & 0.000015 &  1.685331 & 0.003789 & \\
$g0_{2}$  & 0.057852 & 0.000003 & 10.328896 & 0.003628 & \\
$g0_{3}$  & 0.112008 & 0.000005 &  5.158962 & 0.003404 & \\
$g0_{4}$  & 0.139303 & 0.000008 &  3.873643 & 0.003739 & \\
$g0_{5}$  & 0.196875 & 0.000011 &  2.688889 & 0.003788
           & $3g0_3-g0_4$ \\
$g0_{6}$  & 0.219283 & 0.000010 &  3.231557 & 0.003833
           &  $g1_{12}-g1_{7}$\\
$g0_{7}$  & 0.245201 & 0.000014 &  2.040817 & 0.003632
           &  \\
$g0_{8}$  & 0.274591 & 0.000016 &  2.121902 & 0.003754
           & $g1_{14}-g1_6$ \\
$g0_{9}$  & 0.296582 & 0.000019 &  2.147909 & 0.003779 & \\
$g0_{10}$ & 0.315650 & 0.000012 &  3.174317 & 0.003705 & \\
$g0_{11}$ & 0.371213 & 0.000011 &  2.888161 & 0.003661 
          & $g0_2+g0_{10}$\\
$g0_{12}$ & 0.402313 & 0.000014 &  1.955912 & 0.003691
           &  \\
$g0_{13}$ & 0.430279 & 0.000016 &  1.933469 & 0.003667
           &  $2g0_2+g0_{10}$\\
$g0_{14}$ & 0.477989 & 0.000019 &  1.523685 & 0.003714
          & $2g0_{9}-g0_2$\\
$g0_{15}$ & 0.632463 & 0.000022 &  1.142509 & 0.003572
           &  $3g0_3+g0_9$\\
$f0$      & 0.891314 & 0.000039 &  0.610999 & 0.003370
           &  $3g0_{12}-g0_{10}$ \\
$f1$      & 1.226561 & 0.000054 &  0.440674 & 0.003357 
            & $g1_1-3g0_7$ \\
$g1_{1}$  & 1.962055 & 0.000006 &  3.971374 & 0.003391 & \\
$g1_{2}$  & 2.138473 & 0.000023 &  1.117216 & 0.003491
           & $3g0_2+g1_1$ \\
$g1_{3}$  & 2.173656 & 0.000027 &  0.923906 & 0.003474
           &  $g1_{12}-g0_{12}$\\
$g1_{4}$  & 2.212520 & 0.000009 &  2.841390 & 0.003818 & \\
$g1_{5}$  & 2.278367 & 0.000005 &  6.007756 & 0.003646
           &  $g1_6-g0_2$ \\
$g1_{6}$  & 2.336542 & 0.000007 &  3.554191 & 0.003506 & \\
$g1_{7}$  & 2.358879 & 0.000020 &  1.455452 & 0.003754 & \\
$g1_{8}$  & 2.390903 & 0.000013 &  2.152668 & 0.003639 & \\
$g1_{9}$  & 2.430207 & 0.000025 &  1.106695 & 0.003711
           & $3g0_{1}+g1_{8}$\\
$g1_{10}$ & 2.459695 & 0.000036 &  0.745968 & 0.003564
           & $2g1_6-g1_4$ \\
$g1_{11}$ & 2.522384 & 0.000027 &  1.010086 & 0.003635
           &  \\
$g1_{12}$ & 2.578433 & 0.000031 &  1.094429 & 0.003625 & \\
$g1_{13}$ & 2.598238 & 0.000033 &  1.133767 & 0.003647
           & $g1_{14}-g0_{1}$ \\
$g1_{14}$ & 2.612865 & 0.000036 &  0.814589 & 0.003619 & \\
$g1_{15}$ & 2.635155 & 0.000030 &  0.919697 & 0.003693 
          &  $2g1_{12}-g1_{11}$\\
$g1_{16}$ & 2.776246 & 0.000067 &  0.365979 & 0.003447
          & $3g0_4+g1_{7}$   \\
$f2$ & 7.936386 & 0.000056 &  0.427891 & 0.004113 & \\
$f3$ & 9.898140 & 0.000081 &  0.292558 & 0.004691
           &  $g1_1+f2$\\
\hline
\end{tabular}
\end{center}
\end{table}

\bsp	
\label{lastpage}
\end{document}